\documentclass{jfm}
\usepackage{graphicx}
\usepackage{epstopdf, epsfig}
\usepackage{hyperref}
\usepackage{tabularx}
\usepackage{lineno}
\newcounter{magicrownumbers}
\newcommand\rownumber{\stepcounter{magicrownumbers}\arabic{magicrownumbers}}

\shorttitle{Turbulent Momentum Barriers}
\shortauthor{N.O. Aksamit, and G. Haller}

\title{Objective Momentum Barriers in Wall Turbulence}

\author{Nikolas O. Aksamit\aff{1,2}
  \corresp{\email{nik.aksamit@canterbury.ac.nz}},
  George Haller\aff{1}}

\affiliation{\aff{1}Institute for Mechanical Systems, Swiss Federal Institute of Technology,
Zurich, Switzerland
\aff{2}School of Earth and Environment, University of Canterbury,
Christchurch, New Zealand}

\begin{document}

\maketitle

\begin{abstract}

We use the recent frame-indifferent theory of diffusive momentum transport
to identify internal barriers in wall-bounded turbulence. Formed by
the invariant manifolds of the Laplacian of the velocity field, the
barriers block the viscous part of the instantaneous momentum flux
in the flow. We employ the level sets of single-trajectory Lagrangian
diagnostic tools, the trajectory rotation average and trajectory stretching exponent, to approximate both
vortical and internal wall-parallel momentum transport barrier (MTB) interfaces.
These interfaces provide frame-indifferent alternatives to classic velocity-gradient-based vortices and high-shear boundaries
between uniform momentum zones (UMZs). Indeed, we find that these elliptic manifold approximations and MTBs outperform standard vortices and UMZ interfaces in blocking diffusive momentum
transport, suggesting our momentum barriers are physical features that may be the cause of coherence signatures in statistical and non-objective diagnostics. We also introduce normalized trajectory metrics that provide unprecedented visualizations of objective coherent structures by avoiding strong turbulence biases.

\end{abstract}

\section{Introduction}

\begin{figure}
  \centerline{\includegraphics[scale=0.2]{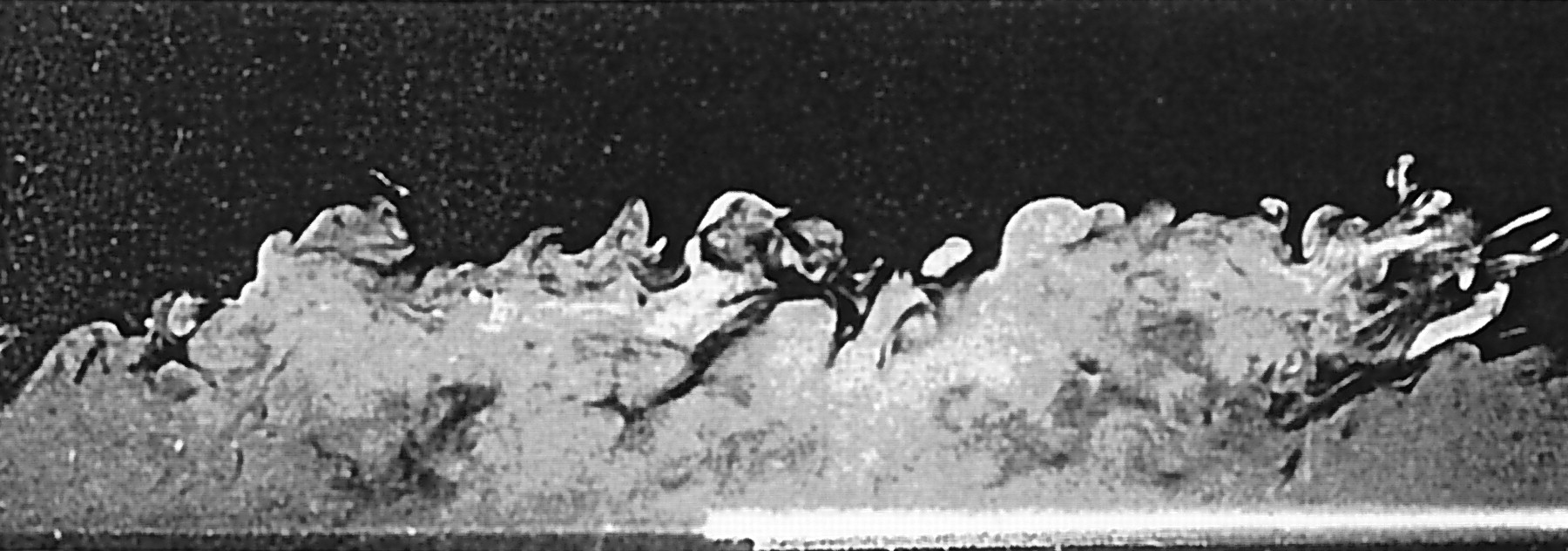}}
  \caption{Material wall-bounded turbulent structures visualized with smoke. Reproduced from \citet{Falco1977} with the permission of AIP Publishing.}
\label{fig:Falco}
\end{figure}

Early studies of turbulent boundary layer structures were fundamentally
inspired by experimentally discovered structures, such as the streaks
in boundary layers photographed by \citet{Kline1967} and the typical eddies
forming large scale motions described by \citet{Falco1977}, shown
in Fig. \ref{fig:Falco}. The pioneering bubble, fog and smoke experiments of the 1960's to 1980's \citep[see, e.g.][]{Fiedler1966, Offen1974, Kline1967, Falco1977, Bandyopadhyay1980, Head1981} quantified intermittent material features and boundary layer structures which coherent structure identification methods continue to rely for validation. 

Boundary layer tracer experiments suggest an organization of fluid by individual vortices at low Reynolds number, and by packets or collections of vortices as turbulence increases. The $33^\circ-45^\circ$ angle at which individual vortices extend from the lower boundary has been recreated in multiple experiments for a variety of Reynolds numbers \citep{Falco1974, Bandyopadhyay1980, Head1981}. The diameter of vortex heads has been found to be a fraction of boundary layer height $\delta$ and scales inversely with Re from $0.01\delta-0.2\delta$ \citep{Falco1974, Falco1977}. Typical vortices also appear to self-organize into bulges, or large scale motions, with streamwise extent ranging from $1.5-2.5\delta$ \citep{Falco1977}, which collectively rise at an inclination angle of $18^\circ - 20^\circ$ \citep{Bandyopadhyay1980, Head1981}. In subsequent hot-wire and particle image velocimetry studies, these ranges of values have been recreated and expanded, but the original measures of material structures remain as the ground-truth for comparison \citep{Adrian2007}.

Advances in experimental techniques and simulation provided highly resolved
velocity fields that stimulated the development of quantitative criteria
for the identification of structures seen in tracer experiments \citep{Adrian2007}.
Some of these criteria extract isosurfaces of velocity components
to define uniform momentum zones (UMZs), while others employ diagnostic
scalar fields, such as the $Q$-, $\lambda_{2}$-, $\Delta$- and
$\lambda_{ci}$-parameters, to define vortices \citep{Adrian2000, Hunt1988, Jeong1995,  Zhou1999, Gao2011}.
Yet other approaches identify a relevant temporal or spatial scale
and employ a conditional averaging prior to feature extraction \citep[see][]{Dennis2011, Gul2020}.

UMZs were first documented in the early experimental study of \citet{Meinhart1995},
leading to the seminal work of \citet{Adrian2000} who suggested that
wall-bounded turbulence may be described as a collection of layered
zonal structures distinguished by their common streamwise velocities. These structures appear between rapid changes in streamwise velocity across strong shear regions where spanwise vorticity may be concentrated. The relevance of these jumps has been supported by analogous jumps in streamwise velocity at turbulent/non-turbulent interfaces and by nearby hairpin vortices and other vortical features. The core statistical methods
used to make these inferences about UMZs and their boundaries were
developed by \citet{Adrian2000} with additional modifications
proposed by \citet{DeSilva2015, Eisma2015, Laskari2018, Fan2019} and others. With
the help of these tools, UMZs have been investigated in a number of
wall-bounded flows, including turbulent boundary layers, channel flows
and pipe flows at various Reynolds numbers \citep{Adrian2000, Kwon2014, DeSilva2015, DeSilva2017, Gul2020}.
UMZs have also been widely used as tools to validate other models \citep{Saxton-Fox2017, Bautista2019}
and generate descriptions of wall-turbulence organization \citep{Adrian2007, Hwang2018}.

Despite these advances, several practical issues are known with the
currently used probability-based UMZ identification scheme, including its sensitivity
to the size of the domain, to the number of velocity vectors used
in identifying significant velocities (and thus extension to three-dimensional data), and to the number of bins used
to reveal streamwise velocity peaks. Two more fundamental issues,
however, also arise. First, the broadly hypothesized role of UMZs
as fundamental turbulence-organizing zonal-like structures that influence the wall-transverse transport of streamwise linear momentum
is often cited but has not yet been directly verified \citep[e.g.,][]{Westerweel2009, Eisma2015, DeSilva2017, Fan2019}. While it is common for researchers to discuss these zonal-structures, and their probable physical importance, there is currently limited understanding of the momentum transport they induce \citep{Montemuro2020}. Nibbling, engulfment, and entrainment are often studied as influential processes for mass and momentum flux, but fluxes through UMZ interfaces are typically estimated indirectly through conditional sampling and average Reynolds stress profiles.
Second, the topology of streamwise velocity level sets (UMZ interfaces)
is not objective, i.e., depends on the frame of reference of the observer.
This is at odds with the tracer patterns arising in foundational tracer
experiments \citep[e.g.,][]{Kline1967, Falco1977, Head1981} that originally inspired
the study of UMZs, given that those patterns are material
and hence are inherently frame-indifferent. Additionally, level-set approaches that rely on Reynolds decompositions, including those in quadrant and octant analysis, introduce a non-physical distortion of the reference frame when different averages are subtracted from each grid cell \citep{Adrian2000b, Kwon2016, Saxton-Fox2017}.

One might dismiss the concern about observer dependence by saying
that a correct understanding of UMZs in the frame of the experiment
is sufficient. The problem with this argument is that a description
of features tied to material observations cannot be correct if it
only holds in the current frame of observation, whether or not one
ever intends to change that frame. More broadly speaking, truly unsteady
flows may have convenient frames but have no distinguished frames,
as already noted by \cite{Lugt1979}. This is the reason why objectivity
(or indifference to Euclidean coordinate changes) as a litmus test for flow feature identification
was already proposed in the 1970's \citep{Drouot1976,Drouot1976a,Astarita1979,Lugt1979},
prompting a number of recent approaches to adopt observer-indifference
as a minimal requirement in coherent structure detection \citep[see][for reviews]{Haller2005,Haller2015,Peacock2015,Kirwan2016,Gunther2018}.

Descriptions of experimentally observed vortical features within and surrounding UMZs \citep[e.g.][]{Head1981}
also face objectivity as a minimal self-consistency requirement. Yet
the $Q$-, $\lambda_{2}$-, $\Delta$- and $\lambda_{ci}$-isosurfaces
used for this purpose are not objective and hence their predictions
for observed material tracer patterns, or momentum transport blocking, cannot be accurate \citep[see][]{Haller2005,Haller2021a}. While correlations may exist between measured scalar concentrations and the features of these non-objective diagnostics \citep[e.g.][]{Westerweel2009, Eisma2021}, there can be no causal relationship.
Several formal modifications of these vortex diagnostics have been proposed
to make them objective, but only the approach of \cite{Liu2019b,Liu2019a}
would be generally applicable, as found by \cite{Haller2021a}. Yet,
for lack of a direct connection to material mixing and transport,
even correct objectivizations of the currently used $Q$-, $\lambda_{2}$-,
$\Delta$- and $\lambda_{ci}$-procedures for vortex identification
would depend on their users. Indeed, the users of these procedures
are expected to pick values for visualized isosurfaces based on their
own expectations for the results \citep[see, e.g.,][]{Dubief2000}.
This commonly used approach results in a subjective
view of the flow, as recently highlighted by \citet{Dong2020}. 

In a parallel development, objective mathematical descriptions of
long-term and short-term material deformation have lead to the notions
of Lagrangian coherent structures (or LCS) and
objective Eulerian coherent structures (or OECS) \citep[see, e.g.,][]{Haller2015, Serra2017, Beron-Vera2018, Serra2020}. Some of these approaches have been used
to identify vortices away from turbulent/non-turbulent interfaces
(TNTIs) in gravity current experiments \citep{Neamtu-Halic2019}, as well as to look at boundary layer structures in PIV studies \citep{Green2007, Pan2009, Wilson2013, He2016, Eisma2021}.
LCSs and OECSs are, however, constructed as boundaries of coherent
structures in passive tracer advection rather than minimizers of
momentum transport, as would be required for a physical UMZ interface analog.

Recently, \citet{Haller2020} developed a theory of objective material
barriers for the transport of active vector fields, such as vorticity
and momentum. These active barriers are sought as material surfaces
that block an objectively-defined transport of momentum or vorticity
more than any other neighboring material surface. Solving this
optimization problem leads to an associated steady, three-dimensional
(3D), incompressible dynamical system (the barrier equation) whose
structurally stable stream surfaces (invariant manifolds) are precisely
the active transport barriers. \citet{Haller2020} showed how active
versions of LCS diagnostics, such as the active finite-time Lyapunov
exponents (aFTLE) and the active polar rotation angle (aPRA) provide
previously unseen levels of detail for momentum-transport barriers
in direct numerical simulations of a turbulent channel flow. Instantaneous
limits of active material barriers can also be extract via the same
machinery. The latter Eulerian barriers are objectively defined surfaces
that block the instantaneous flux of the active vector field in question. 

Here, we use this recent theory of instantaneous active barriers to
define and visualize both momentum trapping vortices and momentum blocking internal interfaces (MTBs) objectively based on their broadly
envisioned role as minimizers of momentum transport. To adapt the active barrier field methods to large datasets, and numerically identify large invariant
manifolds of the active barrier equations, we use recently developed single-trajectory-based
objective coherent structure diagnostics, the trajectory rotation average
(TRA) and trajectory stretching exponent (TSE), from \cite{Haller2021}. Combining these theories,
we develop a simple, systematic procedure that visualizes both MTBs
and momentum-trapping vortices in general 3D, wall-bounded
turbulence. We also show that the active-barrier-based
approach developed here locates vortices and MTB interfaces with significantly
lower viscous momentum flux than nearby surfaces obtained from the broadly used velocity-gradient-based vortex diagnostics and
non-objective UMZ definition.

\section{Methods}

\label{sec:methods} 

\subsection{Objective instantaneous barriers to momentum transport}

For a 3D fluid velocity field $\mathbf{v}(\mathbf{x},t)$ with density
$\rho(\mathbf{x},t)$, the equation of motion can be written as

\begin{equation}
\rho\frac{D\mathbf{v}}{Dt}=-\nabla p+\nabla\cdot\mathbf{T}_{vis}+\mathbf{q},\label{eq:eq of motion}
\end{equation}
where $\frac{D}{Dt}$ is the material derivative, $p(\mathbf{x},t)$
is the equilibrium pressure, $\mathbf{T}_{vis}(\mathbf{x},t)$ is
the viscous stress tensor, and $\mathbf{q}(\mathbf{x},t)$ contains
the external body forces. Fluid trajectories generated by the velocity
field \textbf{$\mathbf{v}$ }are solutions, $\mathbf{x}(t;t_{0},\mathbf{x}_{0})$,
of the ordinary differential equation $\dot{\mathbf{x}}=\mathbf{v}(\mathbf{x},t)$
with initial position $\mathbf{x}_{0}$ at the initial time $t_{0}$.
These fluid trajectories enable the definition of the flow map $\mathbf{F}_{t_{0}}^{t}:\mathbf{x}_{0}\mapsto\mathbf{x}(t;t_{0},\mathbf{x}_{0})$.
A material surface $\mathcal{M}(t)$ is then a two-dimensional (2D)
manifold, 

\begin{equation}
\mathcal{M}(t)=\mathbf{F}_{t_{0}}^{t}[\mathcal{M}(t_{0})],\label{eq:mat surface}
\end{equation}
 evolving under the flow map from its initial position $\mathcal{M}(t_{0})$. 

To identify exceptional momentum-transport minimizing surfaces, we must first agree on a definition of frame-indifferent momentum flux. As pointed out by \citet{Haller2020}, the broadly used linear momentum
flux 

\begin{equation}
\mathrm{Flux}_{\rho\mathbf{v}}(\mathcal{M}(t))=\int_{\mathcal{M}(t)}\rho\mathbf{v}(\mathbf{v}\cdot\mathbf{n})dA\label{eq:old flux}
\end{equation}
is unsuitable for systematic, observer-independent momentum-flux measurements
through a material surface $\mathcal{M}(t)$ for several reasons.
First, this flux expression originally arises from the application
of the Reynold transport theorem to quantify linear momentum carried
by fluid trajectories through a non-material control surface. No such
trajectory crossings are, however, possible through a material surface.
Second, a flux of a quantity through a surface should have the units
of that quantity divided by time and multiplied by the surface area,
which is not the case for $\mathrm{Flux}_{\rho\mathbf{v}}$. Third,
$\mathrm{Flux}_{\rho\mathbf{v}}$ is not objective because under Euclidean
coordinate changes of the form 
\begin{equation}
\mathbf{x}=\mathbf{Q}(t)\mathbf{y}+\mathbf{b}(t),\qquad\mathbf{Q}\mathbf{Q}^{T}=\mathbf{I},\label{eq:Eucliden frame change}
\end{equation}
the integrand in (\ref{eq:old flux}) does not transform as an
objective velocity field, i.e., we have $\mathbf{v}(\mathbf{v}\cdot\mathbf{n})\neq\mathbf{Q}\tilde{\mathbf{v}}(\tilde{\mathbf{v}}\cdot\tilde{\mathbf{n}})$
for the transformed velocity field 
\[
\tilde{\mathbf{v}}=\mathbf{Q}^{T}\left(\mathbf{v}-\dot{\mathbf{Q}}\mathbf{y}-\dot{\mathbf{b}}\right).
\]
This observer dependence is equally true for fluxes obtained from
conditionally-averaged entrainment velocities used in TNTI studies
\citep[e.g.][]{Westerweel2009, DaSilva2014, Eisma2015}, as well as projections of momentum flux in the streamwise direction.

To address these shortcomings of $\mathrm{Flux}_{\rho\mathbf{v}}$,
\citet{Haller2020} introduce a frame-indifferent flux for an arbitrary,
dynamically active vector field $\mathbf{f}(\mathbf{x},t)$ that satisfies
a partial differential equation of the form

\begin{equation}
D\mathbf{f}/Dt=\mathbf{h}_{vis}+\mathbf{h}_{nonvis}\qquad\partial_{\mathbf{T}_{vis}}\mathbf{h}_{vis}\neq0\qquad\partial_{\mathbf{T}_{vis}}\mathbf{h}_{nonvis}=0.\label{eq:evolution equation}
\end{equation}
Here the term $\mathbf{h}_{vis}(\mathbf{x},t,\mathbf{v},\mathbf{f},\mathbf{T}_{vis})$,
arising from diffusive forces (i.e., viscous Cauchy-stresses), is
assumed to be an objective vector field, i.e., $\mathbf{h}_{vis}=\mathbf{Q}\tilde{\mathbf{h}}_{vis}$.
The other term, $\mathbf{h}_{nonvis}(\mathbf{x},t,\mathbf{v},\mathbf{f})$,
is assumed to have no explicit dependence on viscous forces. Instead,
it contains terms originating from the pressure, external forces and
possible inertial effects. For instance, if $\mathbf{f}$ is the linear
momentum of an incompressible Navier--Stokes flow with kinematic
viscosity $\nu$, then the Navier--Stokes equations directly imply
\textbf{
\begin{equation}
\mathbf{h}_{vis}=\rho\nu\Delta\mathbf{v},\label{eq:h_vis for momentum}
\end{equation}
 }which is an objective vector field, because $\Delta\mathbf{v}=\mathbf{Q}\Delta\tilde{\mathbf{v}}$.

The diffusive flux of $\mathbf{f}(\mathbf{x},t)$ through $\mathcal{M}(t)$
can then be defined as the surface integral of the diffusive part
of the surface-normal material derivative of $\mathbf{f}\left(\mathbf{x},t\right)$
over $\mathcal{M}(t)$:
\begin{equation}
\Phi\left(\mathcal{M}(t)\right)=\left[\int_{\mathcal{M}(t)}\frac{D\mathbf{f}}{Dt}\cdot\mathbf{n}\,dA\right]_{vis}=\int_{\mathcal{M}(t)}\mathbf{h}_{vis}\cdot\mathbf{n}\,dA.\label{eq:algebraic flux}
\end{equation}
In contrast to eq. (\ref{eq:old flux}), the diffusive momentum flux
$\Phi\left(\mathcal{M}(t)\right)$ has the correct physical units
of momentum flux and is objective. Indeed, under all observer changes
of the form (\ref{eq:Eucliden frame change}), we obtain $\mathbf{h}_{vis}\cdot\mathbf{n}\,dA=\left(\mathbf{Q}\tilde{\mathbf{h}}_{vis}\right)\cdot\left(\mathbf{Q}\tilde{\mathbf{n}}\right)d\tilde{A}=\tilde{\mathbf{h}}_{vis}\cdot\tilde{\mathbf{n}}\,d\tilde{A}$.

We also define a measure, $\Phi_N$, that is focused solely on quantifying the amount of tangency a surface $\mathcal{M}(t)$ shares with a diffusive momentum flux barrier: 
\begin{equation}
\Phi_N\left(\mathcal{M}(t)\right)=\int_{\mathcal{M}(t)}\frac{\mathbf{h}_{vis}}{||\mathbf{h}_{vis}||}\cdot\mathbf{n}\,dA.\label{eq:algebraic norm flux}
\end{equation}
In flows with a wide range of $\mathbf{h}_{vis}$ magnitudes, $\Phi_N$ provides a normalized measure with no preferential bias towards low-magnitude regions of the flow. 

By formula (\ref{eq:algebraic flux}), a material surface $\mathcal{M}(t)$
is a perfect instantaneous barrier to diffusive momentum flux if $\mathbf{h}_{vis}\cdot\mathbf{n}$ (and thus ${||\mathbf{h}_{vis}||}^{-1}(\mathbf{h}_{vis}\cdot\mathbf{n})$)
vanishes at each point of $\mathcal{M}(t)$. In other words, the surface
$\mathcal{M}(t)$ must be tangent to the vector field $\mathbf{h}_{vis}$
at each of its points, i.e., $\mathcal{M}(t)$ must be an invariant manifold of the
differential equation $\mathbf{x}'=\mathbf{h}_{vis}$ . Here prime
denotes differentiation with respect to the barrier time, $s$, which parameterizes trajectories of this differential equation.

Specifically, when $\mathbf{f}$ is the linear momentum, then we obtain
from \ref{eq:h_vis for momentum} that $\mathcal{M}(t)$ is an invariant
manifold of the instantaneous momentum barrier equation 

\begin{equation}
\mathbf{x}'(s)=\Delta\mathbf{v}(\mathbf{x}(s),t).\label{eq:instant barrier}
\end{equation}
Here, to speed up trajectory integration, we have dropped the small scalar factor $\rho\nu$
in the definition of $\mathbf{h}_{vis}$ in (\ref{eq:h_vis for momentum}). This has no impact on the invariant manifolds (stream surfaces) of $\mathbf{h}_{vis}$.

The barrier time $s$ is a non-dimensional geometric parameter along trajectories of the barrier equations. As such, it has no direct fluid dynamical meaning, much the same way as the geometric parameter $\tau$ has no direct physical meaning in the differential equation $dx/d\tau=v(x,t)$ defining the instantaneous streamlines of a velocity field. However, if we normalize the barrier vector field to a unit vector at each point, then $s$ will measure precisely the arclength of a computed barrier trajectory. In that case, setting a maximal value $s^*$ for $s$ in our calculations will directly control the barrier length scales revealed by the Lagrangian diagnostics computed on the barrier equations up to the barrier time $s^*$.

Even at the Reynolds numbers considered here, the diffusive momentum transport is small relative to the total momentum transport, which is dominated by pressure-induced transport. The ratio between diffusive and pressure-induced momentum transport decreases further with increasing Reynolds numbers. We propose, however, that coherent structure boundaries in the flow are distinguished precisely by their ability to inhibit the diffusive component of the momentum transport. Indeed, the transport induced by the pressure gradient also moves coherent structures along with the bulk flow and hence fails to distinguish their boundaries.  This view on coherent structure boundaries has been justified analytically for all directionally steady Beltrami solutions of the Navier-Stokes equation, for arbitrary high Reynolds numbers \cite[see][]{Haller2020}. The same principle has been verified numerically for diffusive passive scalar transport by \citet{Haller2018}. They find the observed coherent structures in two-dimensional geophysical flows coincide with barriers to the diffusive transport of passive scalars (such as the scalar vorticity) for arbitrarily small diffusion.

As further illustration in the Appendix, we compare perfect barriers to the diffusive momentum transport and perfect barriers to the total momentum transport for an exact Navier-Stokes solution, the time-dependent ABC flow. Due to the dominance of the strongly compressible pressure gradient vector field, the barrier surfaces to total momentum transport accumulate on each other and spiral into fixed points. These characteristically dissipative surfaces are notably dissimilar to the coherent structures seen in tracer experiments and hence would be inconsistent with the view put forward by \citet{Westerweel2009} on the relation between barriers of tracer transport and those of momentum transport.

By construction, any 2D structurally stable invariant manifold $\mathcal{M}(t)$
of the barrier equation (\ref{eq:instant barrier}) represents an exact
and dynamically robust instantaneous barrier to the diffusive transport of linear
momentum. If $\mathbf{v}$ is incompressible, then the barrier equation
(\ref{eq:instant barrier}) is an incompressible, steady dynamical
system, given that the time $t$ only plays the role of a parameter (which temporal frame to investigate)
and the right-hand side of (\ref{eq:instant barrier}) has no explicit
dependence on the barrier time $s$. Therefore, as is well known from
chaotic advection studies of 3D, steady, incompressible flows, structurally
stable 2D invariant manifolds of (\ref{eq:instant barrier}) are stable
manifolds, unstable manifolds and invariant tori. We note that \citet{Haller2020}
also extends the barrier equation (\ref{eq:instant barrier}) to cover
material transport barriers over a finite time interval, but here
we will focus on instantaneous momentum barriers. Both the instantaneous
and the material barrier equations are objective.

Invariant manifolds (or distinguished stream surfaces) of the barrier
equation (\ref{eq:instant barrier}) can only be determined numerically
and hence will be approximate. We provide two different methods to numerically approximate these barriers with different orders of computational burden. In order to evaluate the accuracy of
our computations and compare the momentum-blocking ability of the
computed barriers to nearby features obtained from common vortex and UMZ identification
procedures, we will use the surface-area-normalized geometric momentum
flux across a surface $\mathcal{M}(t)$,

\begin{equation}
\mathrm{\Psi}(\mathcal{M}(t))=\frac{\int_{\mathcal{M}(t)}|\Delta\mathbf{v}\cdot\mathbf{n}|dA}{\int_{\mathcal{M}(t)}dA}.\label{eq:geometric flux}
\end{equation}
This objective quantity does not allow for a cancellation of fluxes
in opposite directions and hence vanishes only on perfectly computed
momentum barriers. As a result, $\mathrm{\Psi}(\mathcal{M}(t))$ provides
an objective, nonnegative scalar metric for the permeability of the
surface $\mathcal{M}(t)$ with respect to momentum transport irrespective
of the size of $\mathcal{M}(t)$.

Similary, the normalized unit barrier field measure,

\begin{equation}
\mathrm{\Psi_N}(\mathcal{M}(t))=\frac{\int_{\mathcal{M}(t)}\left|\frac{\Delta\mathbf{v}}{||\Delta\mathbf{v}||}\cdot\mathbf{n}\right|dA}{\int_{\mathcal{M}(t)}dA},\label{eq:norm geometric flux}
\end{equation}
quantifies the degree of tangency between an imperfect barrier and trajectories of the linear momentum barrier field (\ref{eq:instant barrier}). This measure provides a clear comparison for surfaces in distinct regions of a flow with different momentum barrier field vector magnitudes by focusing solely on geometry with no bias for small $\Delta\mathbf{v}$ values. $\mathrm{\Psi_N}$, however, does not have the units of flux and is thus referred to as our barrier field tangency measure.

\subsection{Identification of momentum barrier surfaces}

A number of relevant techniques have been developed in the LCS literature
to identify distinguished material surfaces of 3D steady flows from
arrays of trajectories \citep[see][for a review]{Haller2015}. These
methods generally require a numerical differentiation of the flow
map or of the velocity field along particle positions. The 3D steady
dynamical system (\ref{eq:instant barrier}) already involves two
spatial derivatives of the velocity field and further spatial
differentiation can only be carried out accurately over sufficiently
dense numerical grids (see \citet{Haller2020} for examples involving
Cauchy-Green strain tensor based diagnostics). 

To avoid the numerical issues associated with further spatial differentiation
of solutions of (\ref{eq:instant barrier}) and to reduce the number of integrated trajectories, here we use very recently developed
single-trajectory diagnostics for elliptic (i.e., vortex-type) and hyperbolic LCS
the trajectory rotation average (TRA) and the trajectory stretching exponent (TSE), derived by \citet{Haller2021}.
On any discretized trajectory $\left\{ \mathbf{x}(s_{i})\right\} _{i=0}^{N}$
of the barrier equation (\ref{eq:instant barrier}) with initial condition
$\mathbf{x}(s_{0})=\mathbf{x}(0)=\mathbf{x}$, the TRA measures the
temporal average of the angular velocity of the trajectory whereas the TSE measures the average hyperbolicity strength along the trajectory. Evaluated
in the context of the barrier equation (\ref{eq:instant barrier}),
these two fields can be computed as

\begin{equation}
\mathrm{TRA}_{0}^{s_{N}}(\mathbf{x})=\frac{1}{s_{N}}\sum_{i=0}^{N-1}\cos^{-1}\frac{\left\langle \dot{\mathbf{x}}(s_{i}),\dot{\mathbf{x}}(s_{i+1})\right\rangle }{\left|\dot{\mathbf{x}}(s_{i})\right|\left|\dot{\mathbf{x}}(s_{i+1})\right|}\label{eq:TRA}
\end{equation}

and 

\begin{equation}
\mathrm{TSE}_{0}^{s_{N}}(\mathbf{x})=\frac{1}{s_{N}}\sum_{i=0}^{N-1}\left|\log\frac{\left|\dot{\mathbf{\mathbf{x}}}(s_{i+1})\right|}{\left|\dot{\mathbf{\mathbf{x}}}(s_{i})\right|}\right|.\label{eq:TSE}
\end{equation}

To simplify our notation, we have omitted the overbar from the $\mathrm{TRA}_{0}^{s_{N}}$ and $\mathrm{TSE}_{0}^{s_{N}}$
which was used by \citet{Haller2021} to distinguish formula (\ref{eq:TRA}) and (\ref{eq:TSE}) 
from their versions that allowed for cancellations along trajectories. TRA and TSE computed in (non-objective) physical velocity fields are
not objective. In our present context, however, $\mathrm{TRA}_{0}^{s_{N}}(\mathbf{x})$ and $\mathrm{TSE}_{0}^{s_{N}}(\mathbf{x})$ are objective fields, because they are computed along trajectories of the objective barrier vector field $\Delta \mathbf{v}$.

Calculating trajectories in the unit barrier field 

\begin{equation}
\mathbf{x}'(s) = \frac{\Delta \mathbf{v}(\mathbf{x}(s),t)}{|\Delta \mathbf{v}(\mathbf{x}(s),t)|}
\label{eq:Unit_Barrier_Field}
\end{equation}
preserves momentum barrier geometry but standardizes the length of all paths for the same advection time, $s_N$. TRA and TSE fields calculated from trajectories in the normalized barrier field, NTRA and NTSE, respectively, visualize features in both highly turbulent regions (large $\Delta \mathbf{v}$), and less turbulent flow regions, with equal fidelity. As will be shown in the following sections, NTRA and NTSE provide unprecedented comparisons of objective coherent structures over the full range of scales and strengths present in a turbulent flow, and concurrently reveal additional weak structures in less turbulent regions that are not evident from other Eulerian methods.

\citet{Haller2021} show that fronts and outer boundaries of nested
cylindrical level surfaces of the TRA and TSE fields highlight the same hyperbolic
and elliptic invariant manifolds as the finite-time Lyapunov exponent (FTLE, \cite{Haller2015}), polar rotation angle (PRA,
\cite{Farazmand2016}) and the Lagrangian-averaged vorticity deviation
(LAVD, \cite{Haller2016a}), but without relying on the spatial differentiation
required by the latter three diagnostics. The TRA and TSE fields are, therefore,
computable from sparse data and their local value is independent of
the number and proximity of other trajectories used in the analysis.
In upcoming visualizations, influential invariant manifolds of the
barrier equation will appear as 2D surfaces along which TRA or TSE exhibit
large changes. The barrier time $s_{N}$ in (\ref{eq:TRA}) and  (\ref{eq:TSE}) can be
selected arbitrarily, as it is independent of the physical time
of the flow data. An increase in $s_{N}$ enhances details in TRA and TSE visualizations, enabling a gradual, scale-dependent exploration
of invariant manifolds in the phase space of the autonomous system
(\ref{eq:instant barrier}).

For arbitrarily large barrier times, however, the quality of visualization begins to degrade. This is caused by barrier field trajectories leaving neighborhoods of  the finite-sized codimension-one invariant manifolds influencing their initial paths. As $\mathrm{TRA}$ and $\mathrm{TSE}$ are monotonically non-decreasing functions of $s$, hyperbolic and elliptic manifolds encountered away from trajectory initial positions will have outsized influence and the diagnostics no longer reflect the features at $\mathbf{x}_0$. We thus suggest determining $s_N$ for a given region $U$ as the decorrelation time of instantaneous $\mathrm{TRA}$ (or  $\mathrm{TSE}$) values. Specifically, for each $\mathbf{x}_0\in U$, we calculate the first zero of the autocorellation

\begin{eqnarray}
R(\tau,\mathbf{x}_0) & = & \sum_{i=0}^{N-1}(f(s_{i},\mathbf{x}_0)-\bar{f})(f(s_{i}-\tau,\mathbf{x}_0)-\bar{f}), \quad \tau\ge0, \nonumber\\ f(s_{i},\mathbf{x}) & = & \mathrm{TRA}_{s_{i}}^{s_{i+1}}(\mathbf{x}_0),
\label{eq:autocorr}
\end{eqnarray}
where $\bar{f}$ is the temporal average of $f$. The median value of this decorrelation time over all $\mathbf{x}_0\in U$ provides a suitable integration time to visualize most invariant manifolds of the barrier field in $U$ with limited interference. We use the same method for determining $s_N$ for the normalized NTRA and NTSE fields in (\ref{eq:Unit_Barrier_Field}) as well. We have found that decorrelation times for $\mathrm{TRA}$ and $\mathrm{TSE}$ to be approximately equal in our numerical studies.

\subsection{Direct numerical simulation data}

To facilitate reproducibility and foster further comparisons with
future developments, we have selected from the publicly available
Johns Hopkins University Turbulence Database (JHTDB) a direct numerical
simulation of a $Re_{\tau}=1000$ channel flow \citep{Perlman2007, Li2008, Graham2016}. On this data set, we compare UMZ interfaces and velocity-gradient-based vortices with
perfect instantaneous barriers to the diffusive transport of linear
momentum. While UMZ studies are typically performed on turbulent boundary
layers, \citet{Kwon2014} and \citet{Fan2019} have argued for the generalization of such
features to channel flows as well. 

The JHTDB channel flow data is available on a $2048\hspace{0.15em}\times\hspace{0.15em}512\hspace{0.15em}\times\hspace{0.15em}1536$
grid for a domain of size $8\pi h\hspace{0.15em}\times\hspace{0.15em}2h\hspace{0.15em}\times\hspace{0.15em}3\pi h$,
where $h$ is the half-channel height. The DNS timestep $\Delta t=0.0013$ in non-dimensional simulation units, with the stored simulation timestep, $\delta t=5\Delta t$, or approximately one channel flow-through time. The analysis herein is conducted over 100 frames spanning the entire simulation database with a duration of 4,000 channel flow-through times from $t=0$ to $t=26$. All figures
and analysis will be displayed in non-dimensional half-channel height
units ($h=1$). This dataset has been used in a number of studies,
most notably by \citet{Bautista2019} to evaluate a model of velocity-based
uniform momentum zones, and by \citet{Jie2021} to investigate inertial particle collection in the quiescent core.

To account for the instability caused by numerical integration through large $\Delta\mathbf{v}$ fluctuations near the channel walls, we have implemented a quadratic buffer in the lower viscous sublayer, within 5 viscous lengths or $0.005h$ from the wall. That is, for $0.995h\le |y|\le1$, we define $\Delta\mathbf{v}(x,y)= v(x,\pm0.995h)(\frac{y\mp1}{0.995h\mp1})^2$ for the appropriate wall. \citet{Eyink2020} examined this region and calculated these heights to be well within the viscous sublayer for our JHU channel. We find that this minimal buffer zone does not modify the visualization of structure in the momentum barrier field, but it significantly expedites calculations and aids in fixed timestep advection of trajectories at the long time scales necessary for determining optimal decorrelation times in (\ref{eq:autocorr}).

\subsection{Computational consideration}

Common to Lagrangian-trajectory diagnostics, the computational burden of calculating TRA and TSE in the vector fields (\ref{eq:instant barrier}) and (\ref{eq:Unit_Barrier_Field}) primarily comes from the accurate integration of trajectories which requires interpolating large vector fields to determine trajectory velocities. The present analysis was conducted in MATLAB on either a professional workstation for specific examples, or a high performance computing cluster to calculate large numbers of TRA and TSE fields in large domains. Using a fourth-order Runge-Kutte integration scheme with 10,000 timesteps on a 2.3 GHz, 18 Core, 128GB iMac Pro, TRA and TSE values from 10,000 initial positions, $\mathbf{x}_0$, can be computed in an interpolated vector field of $100 \times 100 \times 100 $ grid points, with spatial dimensions $1.2 h \times 0.6 h \times 0.6 h$, in a wall-clock time of 5 seconds. This can clearly be improved upon in more computationally efficient programming languages. 

As one increases the spatial dimensions of initial conditions $\mathbf{x}_0$, the flow domain used in the interpolation needs to be increased to avoid boundary effects, as in \citet{Eisma2021}. This, in turn, decreases the speed of the computations. We thus calculate the TRA and TSE fields in (\ref{eq:instant barrier}) using a patchwork of initial positions in flow domains with large margins so as to avoid trajectories leaving the domain. In the subsequent analysis, we found exemplary visualization is possible with spatial resolutions on the order of 1 to 10 viscous lengths ($10^{-3}h$ to $10^{-2} h$), with the finer resolutions beneficial for smooth flux-minimizing isosurface extraction. Parallelized MATLAB scripts to calculate TRA and TSE are available on github (\url{https://github.com/haller-group/TRA_TSE}).

\section{Results}

\label{sec:results}

\subsection{Objective momentum transport barrier visualization}

\begin{figure}
  \centerline{\includegraphics[scale=0.2]{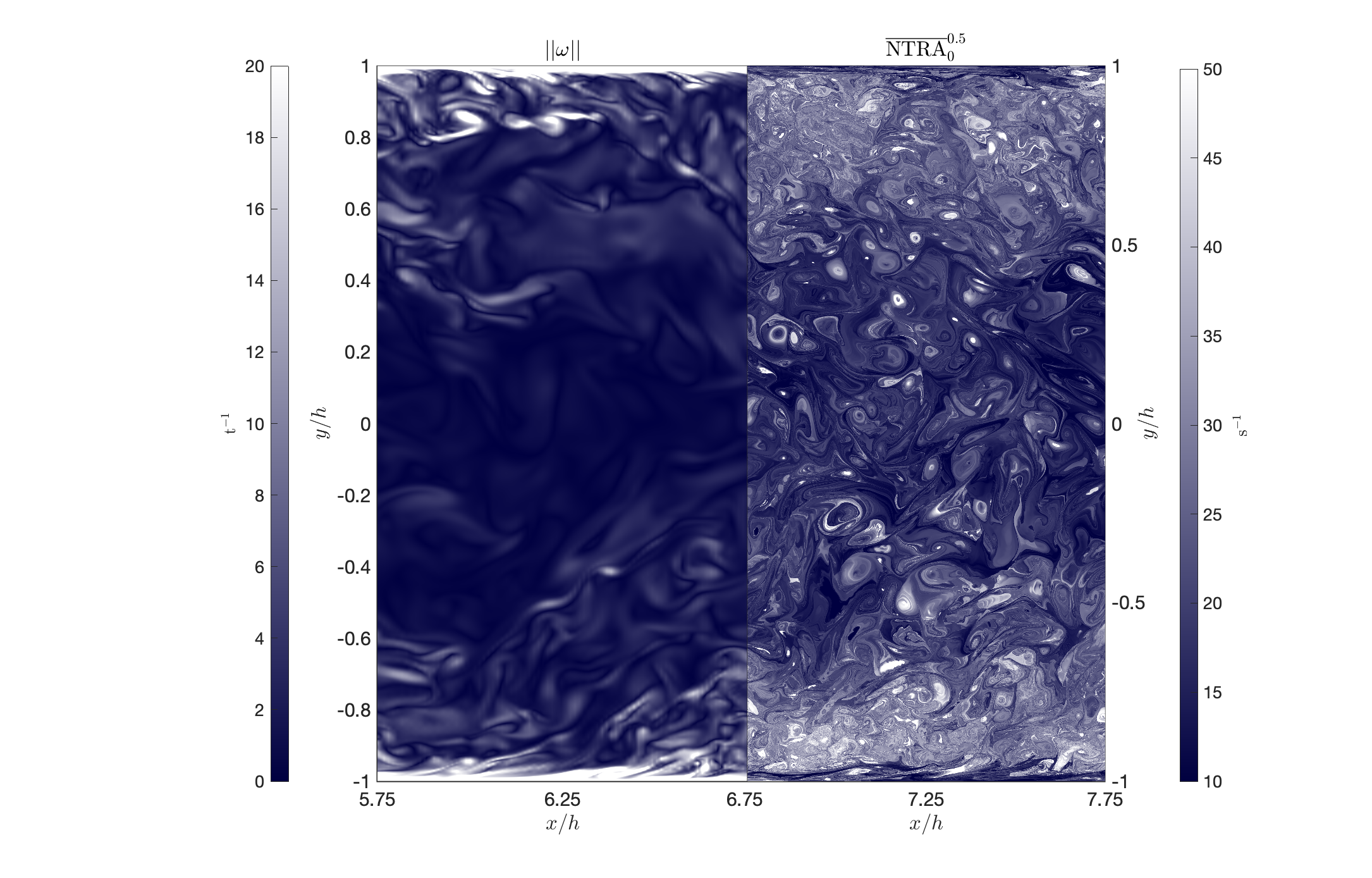}}
 \caption{A streamwise-wall-normal plane of the $Re_{\tau}=1000$ JHTDB Channel Flow colored
by vorticity magnitude (left) and $\mathrm{\overline{NTRA}}_{0}^{0.5}$ (right). Both visualizations of turbulent features were calculated at the same spatial resolution from the same velocity data at one time step. The enhanced visualization possible with $\overline{\mathrm{NTRA}}$ provides a striking comparison with classic techniques and illuminates faint and weak structure in the center of the channel while maintaining objectivity.}
\label{fig:ChannelVis}
\end{figure}

Figure \ref{fig:ChannelVis} compares a common turbulence visualization diagnostics, the vorticity magnitude, with the $\mathrm{NTRA}$ field
 computed for the normalized momentum barrier equation (\ref{eq:Unit_Barrier_Field})
on a streamwise-wall-normal plane of initial conditions in the channel at the non-dimensional DNS time $t=0.065=5\delta t$. The predominant channel
flow is in the positive $x$-direction with channel walls at $y=\pm 1$. Both diagnostic fields in Fig. \ref{fig:ChannelVis} were generated from precisely the same velocity field, and were calculated and visualized at the same spatial resolution ($\delta_x = \delta_y = 10^{-3}$). The vorticity picture is indicative of the scale and resolution of structures that are captured by velocity and velocity-gradient-based level-set methods. The barrier trajectories $\mathbf{x}(s)$ used in these simulations
were advected under the 3D normalized barrier equation until $s_N=0.5$, the order of decorrelation time for the entire channel.

The shear generated by the upper and lower channel walls is evident
from the high rates of trajectory rotation and vorticity. Surprisingly, in the NTRA field, there is strong evidence of many more vortical momentum barriers in the center of the channel, which is typically viewed as a quiescent region. This shows that while there are still many complex vortex interactions occurring in the region, their relatively weaker signature makes them impossible to identify in the weak gradients and uniformly low vorticity values. Thus, the NTRA field provides an enhanced visualization of structures at a much wider range of spatial scales and structure strengths for the same underlying velocity data. For identifying boundaries and the structure extraction discussed in the next sections, we find the large changes evident in TRA and TSE fields to be most beneficial. At the same time, NTRA and NTSE continue to provide homogeneous fidelity visualizations of barriers in large domains that contain a wide range of vector magnitudes.

\begin{figure}
  \centerline{\includegraphics[scale=0.22]{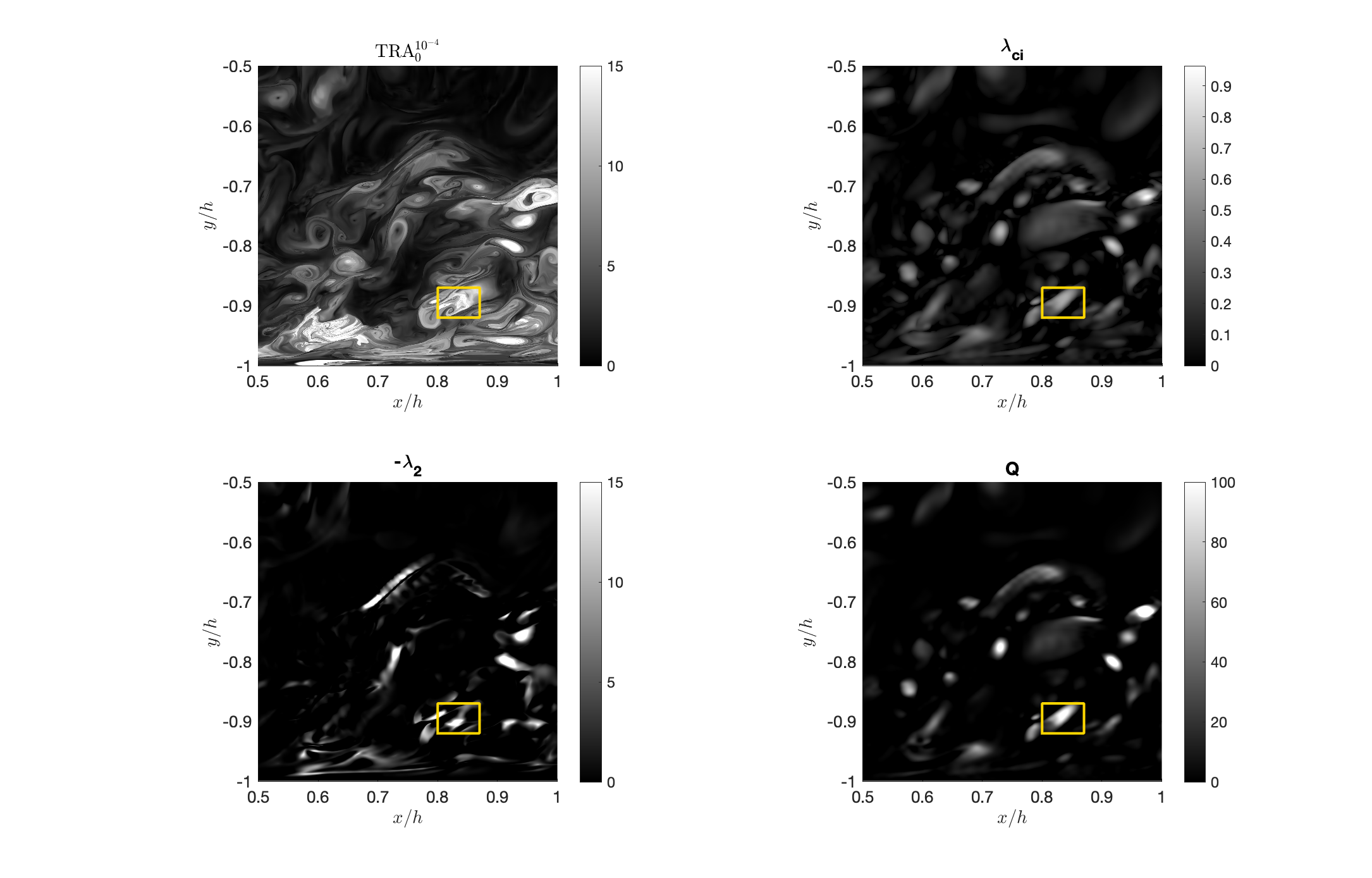}}
  
 \caption{Comparison of the objective TRA field with non-objective, velocity-gradient-based
vortex identification diagnostics in the highly turbulent region at
the channel wall. All four plots were computed with the same spatial
resolution from the same single velocity snapshot of the DNS data.}
\label{fig:FourComp}
\end{figure}

Zooming in on the turbulent wall-region, we find the degree of detail of TRA fields for the original barrier equations (\ref{eq:instant barrier}) is also unattainable by classic velocity-gradient based vortex diagnostics. This first-order benefit can be seen in Fig. \ref{fig:FourComp},
in which TRA is compared with $Q$, $\lambda_{2}$, and $\lambda_{ci}$
(swirling strength) \citep{Hunt1988, Jeong1995,  Zhou1999} for a streamwise-wall-normal ($x$,$y$) plane
adjacent to the lower channel wall. Again, all four plots were generated
with the same spatial resolution from the same single velocity snapshot
of the DNS data. In contrast to the other plots, the TRA is objective and reveals substantially more of the complexity
of the flow for the decorrelation advection time $s_{N}=10^{-4}$.

The TRA plot in Fig.
\ref{fig:FourComp} reveals a complex connection network and a layering
of unique rotational features not present in the velocity-gradient-based
diagnostics. All colormaps have been chosen to reveal the full range of metric values, though gradient-based approaches suffer from the same issues as in  Figure \ref{fig:ChannelVis}. The level of detail in the TRA field provides increased accuracy in vortex
detection. For example, at approximately $(x/h,y/h)=(0.8,-0.9)$, there
is a clear maximum in all three velocity-gradient-based metrics, suggesting
the potential presence of a coherent vortex (yellow box). Upon closer inspection of the TRA
in the same region, however, we find a lack of nested cylindrical
TRA level surfaces. Instead, filamenting invariant surfaces of the
barrier equation are present that are not structurally stable and
hence do not define vortical barriers to momentum transport. The TRA field discerns these important structural features and has a significant advantage in preventing false-positive vortex identifications. Furthermore, these details are useful for accurately tracking individual features from one time to the next, as will be discussed in Section \ref{sec:Statistics}.
An internal layering close to the wall is also present in the TRA
field, as is the organization of vortices around a clearer transition between the more turbulent wall
region and the less turbulent channel core. We discuss this interface
in more detail in the Section \ref{Section:MTB}.

\subsection{Momentum-trapping vortices}
\label{Section:Vortices}

In 2D cross-sections of the flow, vortex boundaries can be located
as outermost members of nested families of closed level curves of
the TRA. Launching trajectories of the 3D barrier equation (\ref{eq:instant barrier})
from these boundary curves generates instantaneous, vortical momentum
barrier surfaces in the full 3D flow. In direct contrast to velocity-gradient-based
vortex identification practices, this process is devoid of any user-defined
parameters beyond a choice of spatial resolution of barrier-field trajectory initial positions, which only serves to control the level of detail
in the TRA field. In contrast to velocity and velocity-gradient-based diagnostics, increasing the spatial resolution of TRA and TSE fields beyond that of the underlying velocity field can continue to increase the structural information revealed because of the dummy time barrier field integration.

\begin{figure}
  \centerline{\includegraphics[scale=0.2]{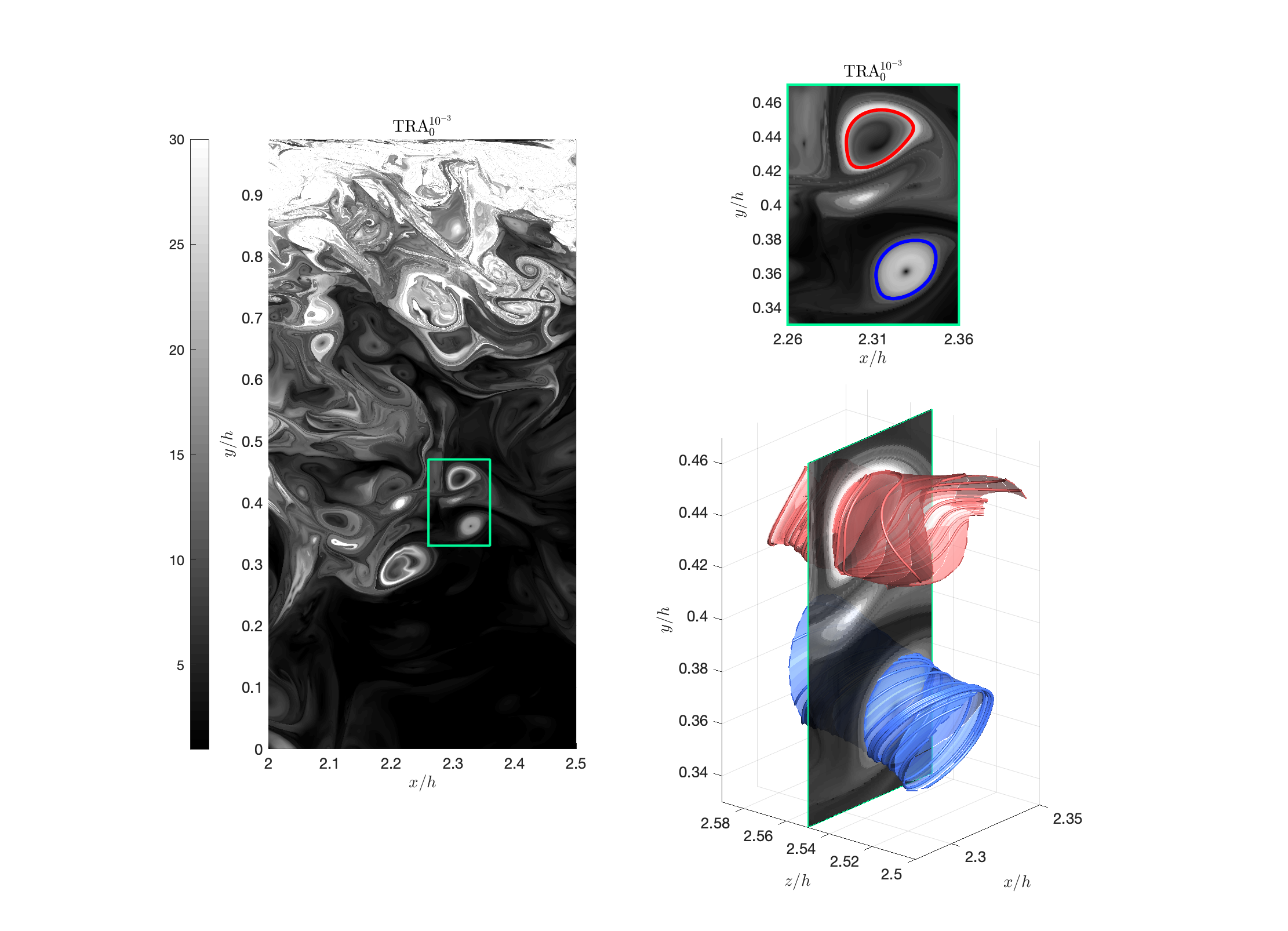}}
 \caption{Momentum-trapping vortices in the turbulent channel flow. The vortex
boundaries are determined as streamsurfaces of $\Delta\mathbf{v}$
that intersect the plane of investigation ($z=2.55$) along outermost
closed and convex TRA contours.}
\label{fig:MtmCore}
\end{figure}

Figure \ref{fig:MtmCore} shows one example of our momentum-trapping
vortex identification method, with barrier trajectories starting from the
$z=2.55h$ plane. The left panel also reveals a strong similarity between the detailed
structures in the TRA field and the material boundary layer structures
visualized in smoke experiments (Fig. \ref{fig:Falco}), to be discussed in more detail in Section \ref{sec:Statistics}.
 Through a simple search algorithm on TRA contours, we have identified
a region with a nested set of closed TRA level curves. By extracting all TRA contours that that span the range of values present, we can isolate the closed and convex contours. The outermost
convex boundary curve for each vortex obtained in this fashion is
highlighted in the top right of Fig. \ref{fig:MtmCore}. Using these vortex boundaries as initial positions, we can advect a dense set of trajectories in the barrier field in forward and backward barrier time to obtain the exact momentum transport barriers seen in the bottom right. If choosing
smaller members of the set of closed level curves of the TRA as a curve of
initial conditions, we obtain an internal foliation of the vortex
by smaller cylindrical momentum barriers. These block radial diffusive
momentum transport within the vortex. The details of this process are described in Algorithm 1.

\begin{table}
\def~{\hphantom{0}}
\begin{tabularx}{\linewidth}{|l|X|}
{} & {\bf Algorithm 1: Extracting momentum vortex cores} \\
\rownumber & For a co-dimension 0 domain of interest $U\subset \mathbb{R}^3$, select a grid of initial conditions $\mathbf{x}_0\in U$. \\
      \rownumber & Choose an initial barrier time estimate $\epsilon$ and calculate the discretized trajectories $\left\{ \mathbf{x}(s_{i})\right\} _{i=0}^{N}$ in the barrier field (\ref{eq:instant barrier}) where $\epsilon=s_N$.\\
      \rownumber & For each trajectory $\mathbf{x}(s_{i}) _{i=0}^{N}$, calculate  $f(s_{i},\mathbf{x})$ following (\ref{eq:autocorr}) and (\ref{eq:TRA}).\\
      \rownumber & Determine the optimal barrier time $\epsilon_0$ as the first-zero crossing of the autocorrelation of $f$ as defined in (\ref{eq:autocorr}). If no zero-crossing exists, repeat Steps 2-4 with progressively larger $\epsilon$ until a zero-crossing exists. \\
      \rownumber & Repeat Steps 1-2 for $\epsilon_0=s_N$ and calculate $\mathrm{TRA}_0^{s_N}$ following (\ref{eq:TRA}).\\
       \rownumber & Select a plane of interest of initial conditions that intersects $U$ and extract iso-contours in that plane for the range of TRA values generated in Step 5. Record only the outermost closed convex contours $\left\{ \gamma_{j}\right\}$ as vortex core candidates. \\
       \rownumber a & {\bf To extract the invariant manifold} of (\ref{eq:instant barrier}) that intersects a given curve $\gamma_j$ , identify trajectories $\left\{ \mathbf{x}_{\gamma_j}(s_{i})\right\} _{i=0}^{N}$ such that $\mathbf{x}_{\gamma_j}(0)\in\gamma_j$. If an insufficient number of trajectories exist, identify points in $\gamma_j$ through interpolation to obtain a set $\mathbf{x}_{\gamma_j}(0)$ and repeat Step 2 with $s_N=\epsilon_0$. \\
      7b & Given a sufficiently dense set of trajectories, the co-dimension 1 vortex core be identified by linear interpolation between nearest points in the set of trajectory data $\left\{ \mathbf{x}_{\gamma_j}(s_{i})\right\} _{i=0}^{N} \in \mathbb{R}^3$. \\
            8a & {\bf To obtain a level-surface approximation} of a momentum vortex core that intersects a given closed contour $\gamma_{j}$, extract that TRA isosurface $\mathcal{I}_{\mathrm{TRA}}$ that intersects $\gamma_{j}$. \\ 
      8b & Repeat 7a and calculate the distance between points in $\mathcal{I}_{\mathrm{TRA}}$ and nearest points in $\left\{ \mathbf{x}_{\gamma_j}(s_{i})\right\} _{i=0}^{N}$. \\
     8c & Calculate the PDF of these distances. $\mathcal{I}_{\mathrm{TRA}}$ will be sufficiently close to $\left\{ \mathbf{x}_{\gamma_j}(s_{i})\right\} _{i=0}^{N}$ for many points, resulting in a distinct peak in the PDF near zero. Refine $\mathcal{I}_{\mathrm{TRA}}$ to only include points in this near-zero peak\\ 
     9 & Steps 5-8 can be repeated for $-\epsilon_0$ to further expand the extracted vortex core.
  \end{tabularx}
\label{tab:kd}
\end{table}

A simpler but only approximate way to visualize objective momentum
barriers is to plot level surfaces of the TRA field. This is inspired
by the observation that structurally-stable elliptic invariant manifolds (such
as invariant tori) of the barrier equation (\ref{eq:instant barrier})
will be spanned by trajectories with the same averaged angular velocities
in the limit of $s_{N}\to\infty$. For finite values of $s_{N}$,
this relationship is only approximate and hence TRA isosurfaces are
only proxies to exact invariant manifolds formed by the trajectories
of (\ref{eq:instant barrier}). For such finite values, nearby particle
trajectories that do not lie on the same invariant manifold may also
accumulate the same TRA value. As a consequence, contour-plotting algorithms
may connect approximations of different momentum barriers into one
approximate level surface. Such artifacts arising from this simplified
visualization can be discounted by launching actual barrier trajectories
of (\ref{eq:instant barrier}) from the intersections of TRA level
sets from a reference cross section and discounting parts of the level
surface whose distance from such barrier trajectories exceeds a tolerance
value. 

\begin{figure}
  \centerline{\includegraphics[scale=0.20]{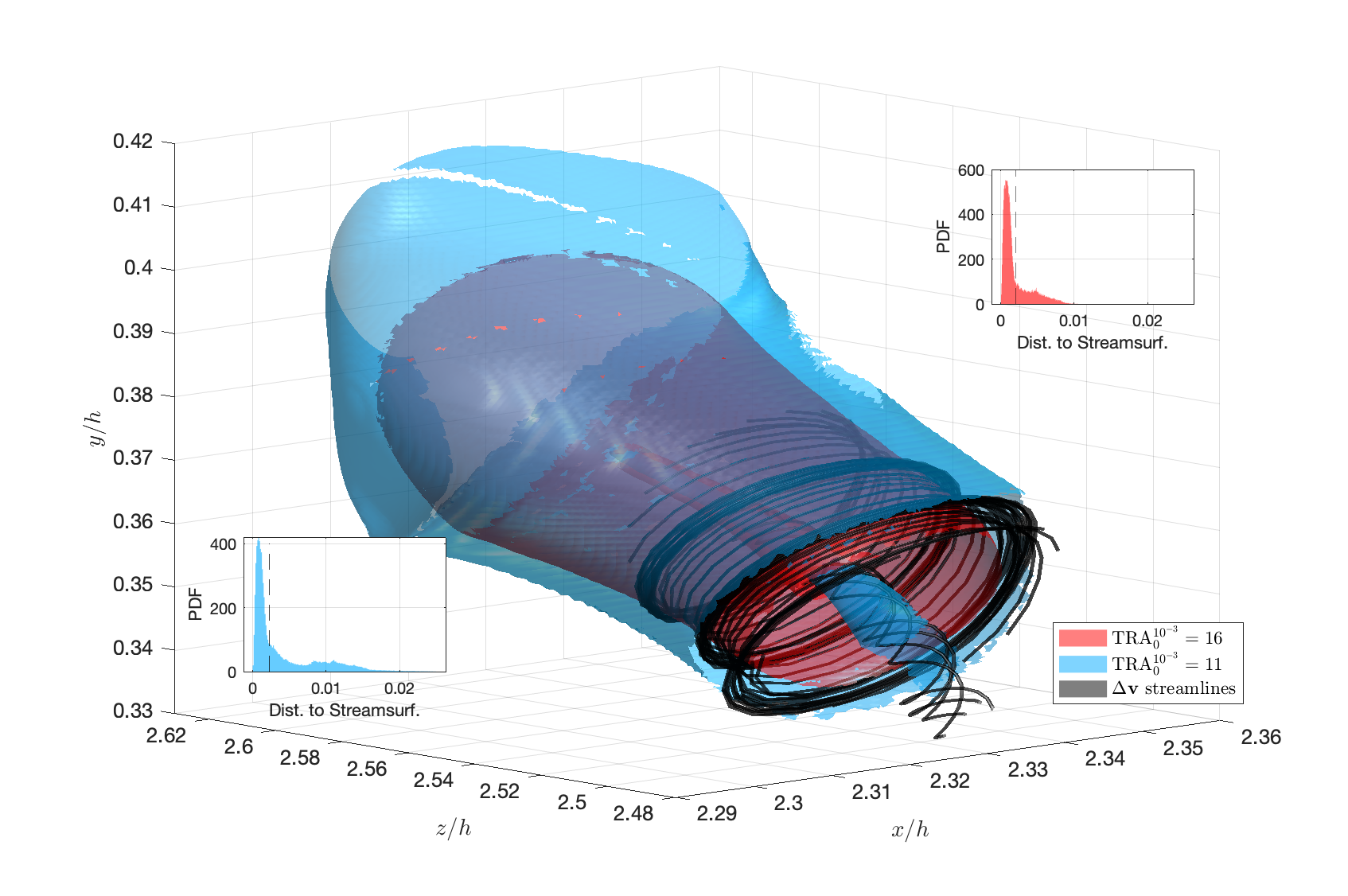}}
 \caption{TRA isosurfaces as approximations to the momentum transport barriers
shown in Figure \ref{fig:MtmCore}. Blue and red surfaces correspond
with two distinct $\mathrm{TRA}_{0}^{10^{-3}}$ values. Shown in black
are a subset of streamlines initiated on the $\mathrm{TRA}_{0}^{10^{-3}}=16$
contour on the $z=2.55$ plane. Inset are probability distribution
functions of the minimum distance of unfiltered surface points to
streamlines initialized on TRA contours at $z=2.55$.}
\label{fig:IsoCompare}
\end{figure}

An example of this isosurface separation process is detailed in Figure
\ref{fig:IsoCompare}. 
Starting with the vortex identified by the blue 2D convex
contour in Figure \ref{fig:MtmCore}, multiple concentric 3D TRA shells
can be seen in Figure \ref{fig:IsoCompare}. The outer and inner blue
shells correspond with $\mathrm{TRA}_{0}^{10^{-3}}=11$ level sets, and
the two red shells correspond to a higher rotation rate, the $\mathrm{TRA}_{0}^{10^{-3}}=16$
level set. Probability distribution functions of the distance between
each isosurface and the barrier field streamlines generated from their
intersection with the $z=2.55h$ plane is shown inset in Figure
\ref{fig:IsoCompare}. As is typical for all vortices we have investigated,
there is a clear probability density function (PDF) peak close to
zero that can be automatically isolated for both isosurfaces with a variety of algorithms, including inflection points or kernel density estimation. These
values correspond with points on the TRA level surface that closely
approximate the momentum-blocking invariant manifolds in Figure \ref{fig:MtmCore}. Once points with
streamsurface-distances outside this peak are removed from the visualization,
the separation between each vortex shell is clearly visible. The selected distances of separation used in this visualization are shown in the inset PDFs as dashed vertical lines. The details of this process are also summarized in Algorithm 1.

To quantify how well outermost cylindrical TRA level surfaces, denoted
$\mathcal{I}_{\mathrm{TRA}}$, approximate true momentum barriers,
we calculate the normalized objective geometric flux $\Psi\left(\mathcal{I}_{\mathrm{TRA}}\right)$
defined in (\ref{eq:geometric flux}), which vanishes only on perfect
barriers to momentum transport. For comparison, we also extract representative
isosurfaces of $\lambda_{ci}$, $\lambda_{2}$ and $Q$, denoted by
$\mathcal{I}_{\mathrm{\lambda_{ci}}}$, $\mathcal{I}_{\mathrm{\lambda_{2}}}$
and $\mathcal{I}_{\mathrm{Q}}$, in the same 3D fluid volume and calculate
$\Psi$ on these surfaces as well. While there are various empirical
values proposed for representative $\mathcal{I}_{\mathrm{\lambda_{ci}}}$,
$\mathcal{I}_{\mathrm{\lambda_{2}}}$ and $\mathcal{I}_{\mathrm{Q}}$
isosurfaces (see, e.g., \cite{Jeong1995,Zhou1999,Ganapathisubramani2006,Gao2011,Dong2020}),
we initially generate surfaces that correspond with their originally
argued value, dividing strain-dominated and rotation-dominated regions.
This value is zero for all velocity-gradient-based metrics
\citep[see][]{Hunt1988, Jeong1995,  Zhou1999}.

\begin{figure}
\centerline{\includegraphics[scale=0.17]{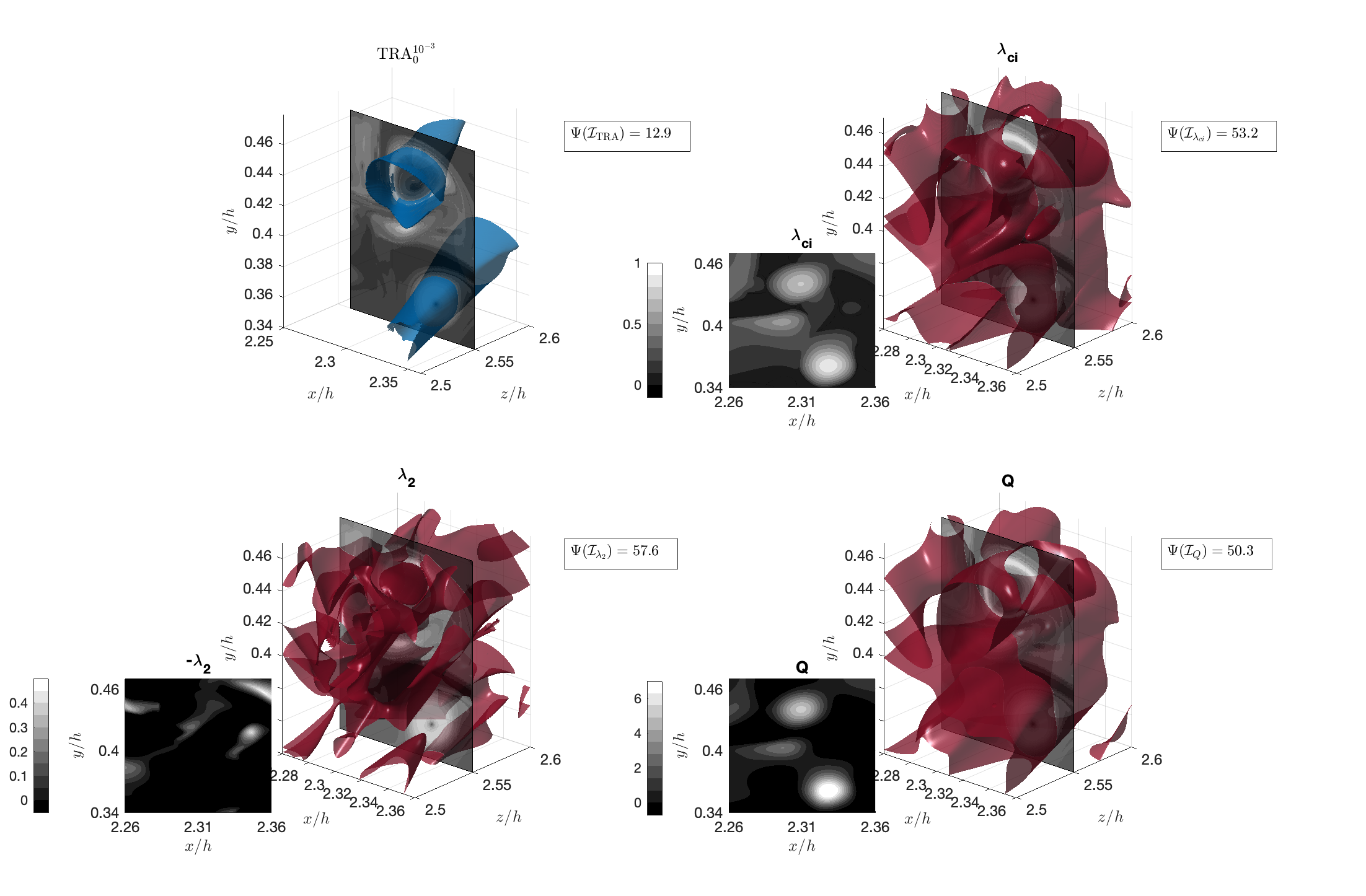}}
 \caption{Comparison of area-normalized objective momentum flux values for vortex
boundaries identified from different vortex identification methods.}
\label{fig:IsoCompare2}
\end{figure}

The isosurfaces generated for each scalar metric are displayed in
Figure \ref{fig:IsoCompare2}. 
As common practice, the swirling strength $\lambda_{ci}$ has been
normalized by its maximum value in the volume of interest. Each isosurface
is shown as it intersects the $z=2.55h$ plane colored by $\mathrm{TRA}_{0}^{10^{-3}}$.
For the three velocity-gradient-based diagnostics, the corresponding
diagnostic field on $z=2.55h$ is shown inset next to the volumes.
The surface-area-normalized momentum flux across each isosurface is
also noted to the right of each respective volume.

In the $z=2.55h$ insets, $\lambda_{ci}$, $\lambda_{2}$ and $Q$
values all show some indication of the presence of the upper vortex
from the TRA field, though evidence of the lower vortex is not present
in $\lambda_{2}$. This suggest the presence of our physical momentum barriers may have influenced these scalar fields, but the metrics are ineffective at predicting where the transport barriers lie. Even as simplified approximations of true linear-momentum
barriers, TRA isosurfaces are still quite effective momentum transport barriers. Indeed, $\Psi(\mathcal{I}_{\mathrm{TRA}})$
is only $24\%$ of $\Psi(\mathcal{I}_{\mathrm{\lambda_{ci}}})$, $22\%$
of $\Psi(\mathcal{I}_{\mathrm{\lambda_{2}}})$ and $26\%$ of $\Psi(\mathcal{I}_{\mathrm{Q}})$. Recall that $\Psi=0$ for the invariant manifolds in Figure \ref{fig:MtmCore} that $\mathcal{I}_{\mathrm{TRA}}$ is approximating here.

We have found that with increased numerical accuracy in our barrier
field integrations, TRA calculations improved and resulted in a further
decrease in $\Psi(\mathcal{I}_{\mathrm{TRA}})$. This indicates one
can more closely approximate the true momentum transport barriers
with more computational expense when a particular region of interest
is identified. We have not found analogous numerical improvements
in flux reduction upon refining the spatial differentiation or resolution
of the velocity-gradient-based methods. Overall, we find that the classic
vortex diagnostics we have tested do not provide a clear indication
of a pair of 3D vortices when one directly implements the criteria
arising from their theoretical derivations. Rather, one must hand pick values to obtain a vortical feature in two or three dimensions as noted by \citet{Dubief2000, Dong2020}. This is problematic in turbulent
flows where there is no ground-truth of structure topology against
which one can validate their hypothesized threshold values.

\begin{figure}
\centerline{\includegraphics[scale=0.15]{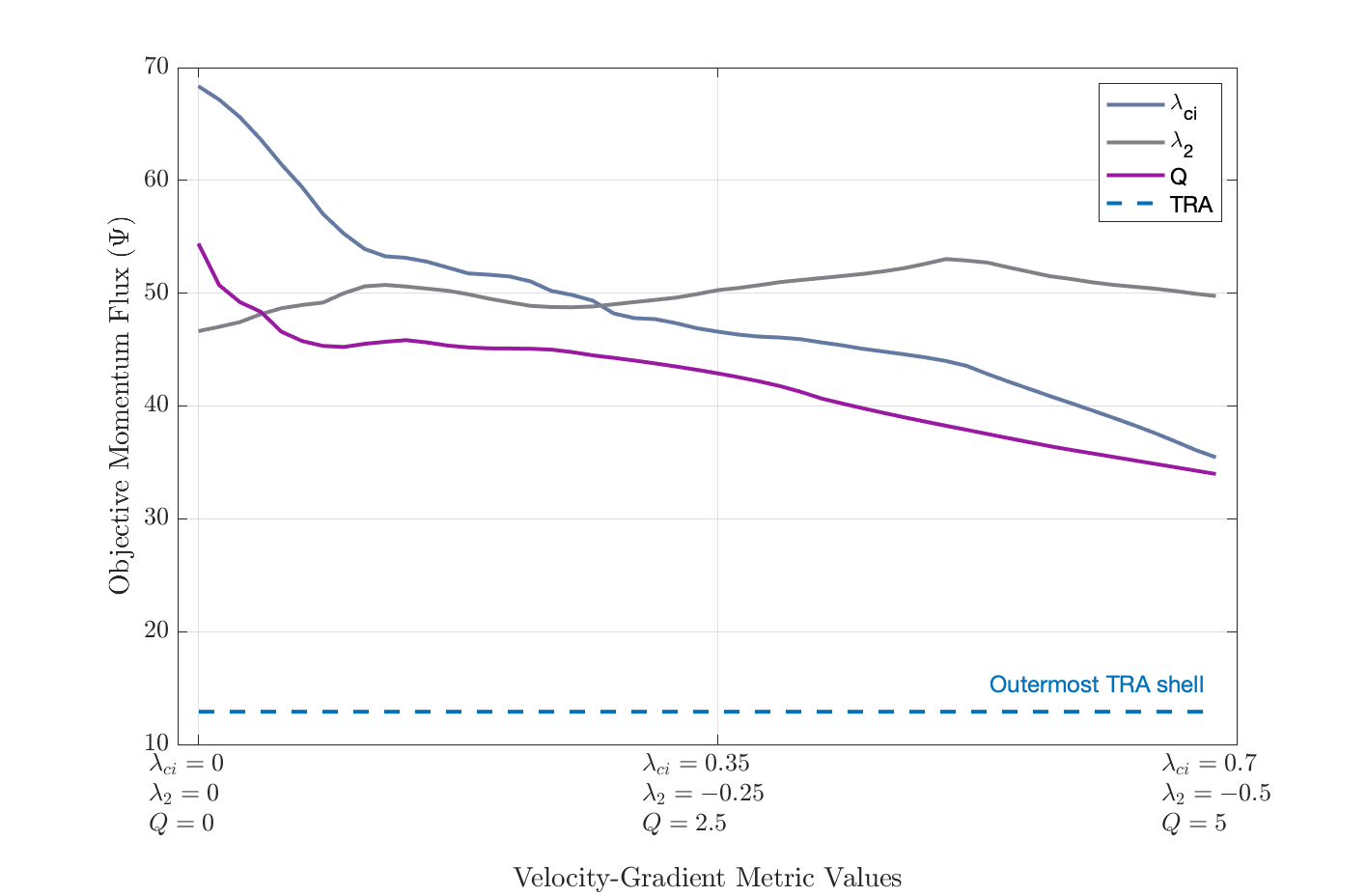}}
 \caption{Comparison of momentum flux through $\mathcal{I}_{\mathrm{TRA}}$
and through structure boundaries defined by a wide range of isosurface
values for the $\lambda_{ci}$, $\lambda_{2}$ and $Q$ metrics.}
\label{fig:FluxCompare}
\end{figure}

To accommodate the wide range of empirical or heuristic thresholds
used in the literature, we have also performed the same flux calculations
for $\lambda_{ci}$, $\lambda_{2}$ and $Q$-isosurfaces for a range
of values that includes and exceeds available suggestions found in
the literature. The resulting fluxes are displayed in Figure \ref{fig:FluxCompare}.
Note that $\mathcal{I}_{\mathrm{TRA}}$ in blue continues to outperform all
other diagnostics over the whole range of empirical threshold values
for the latter diagnostics. 

\begin{figure}
\centerline{\includegraphics[scale=0.25]{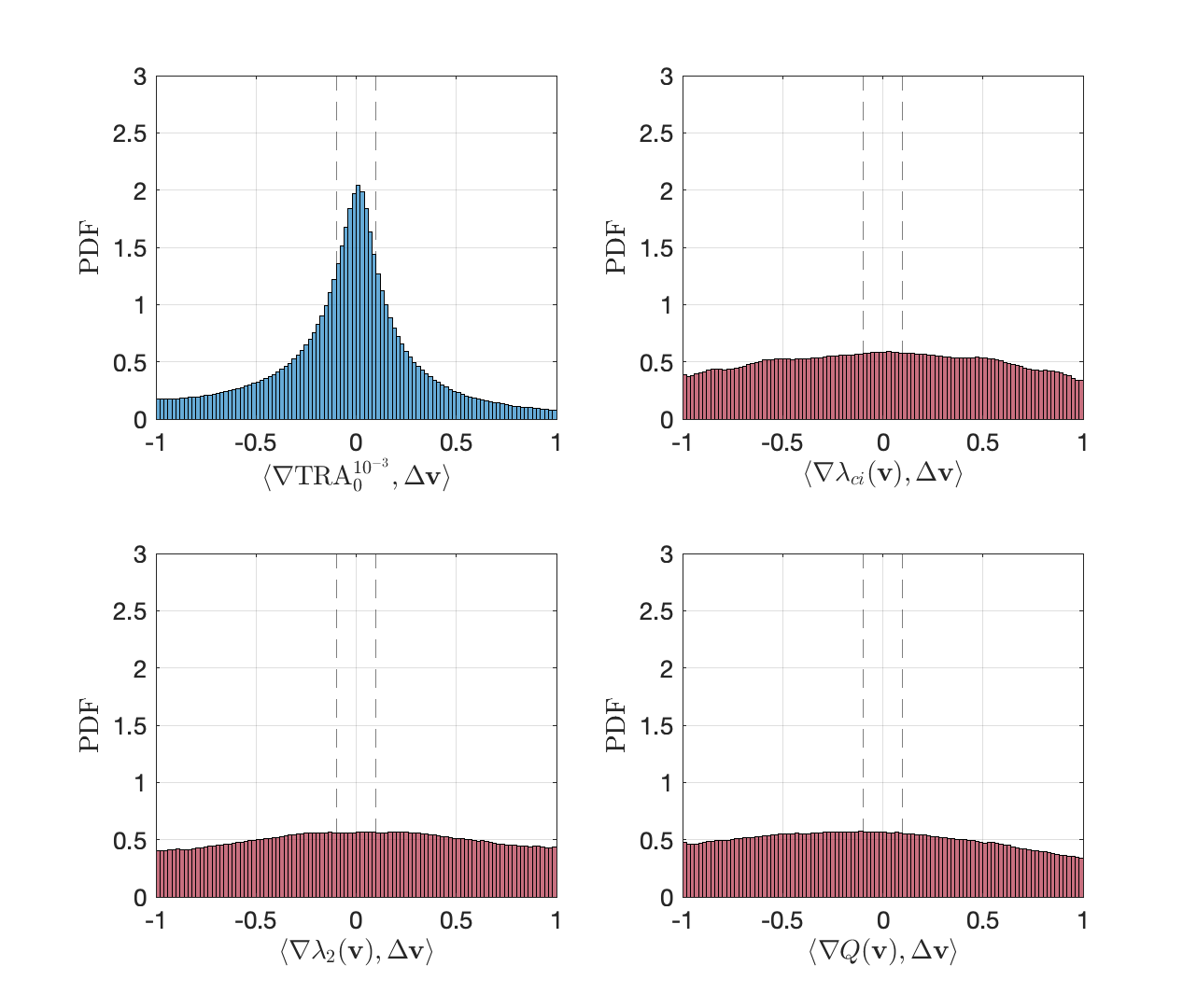}}
 \caption{Probability distributions of the normalized inner product of momentum
barrier field vectors and isosurface normals for TRA and the three
standard velocity-gradient-based diagnostics for a 3D rectangular
volume of fluid containing the vortices in Figure \ref{fig:IsoCompare2}.
The clear singular peak around 0 in the TRA PDF indicates a strong
agreement between TRA surfaces and the underlying momentum transport
blocking interfaces for both elliptic and hyperbolic surfaces. Similar
behavior does not exist for the other three diagnostics. A $\pm5^{\circ}$
difference between surface tangents and barrier vectors is delimited
by dashed lines.}
\label{fig:InnerProd_Vel}
\end{figure}

In Figure \ref{fig:InnerProd_Vel}, we verify the surface tangency of arbitrary-valued $\mathrm{TRA}$ level-surfaces
with invariant manifolds.
If isosurfaces of a given scalar field were exact streamsurfaces of
the barrier equation, then their normals (i.e., the gradient of that
scalar field) would be perpendicular to $\Delta\mathbf{v}$. Each
subplot of Figure \ref{fig:InnerProd_Vel} shows the probability distribution
of inner products of normalized scalar field gradient vectors with
the normalized barrier field, $\Delta\mathbf{v}/\left|\Delta\mathbf{v}\right|$,
over the entire 3D volume containing our vortices ($[2.26,2.36]\times[0.34,0.37]\times[2.5,2.6]$). These are precisely distributions of the signed integrands of the numerator in (\ref{eq:norm geometric flux}).
Vertical dashed lines mark a $\pm5^{\circ}$ deviation from perfect
agreement between momentum barriers and barriers generated by each
diagnostic level set. Notably, there is a clear peak around 0 for
$\boldsymbol{\nabla}\mathrm{TRA}_{0}^{10^{-3}}$ , while a nearly uniform
(random) distribution can be seen for angles between barrier field
vectors and velocity-gradient-based isosurface normals. 

\begin{figure}
\centerline{\includegraphics[scale=0.25]{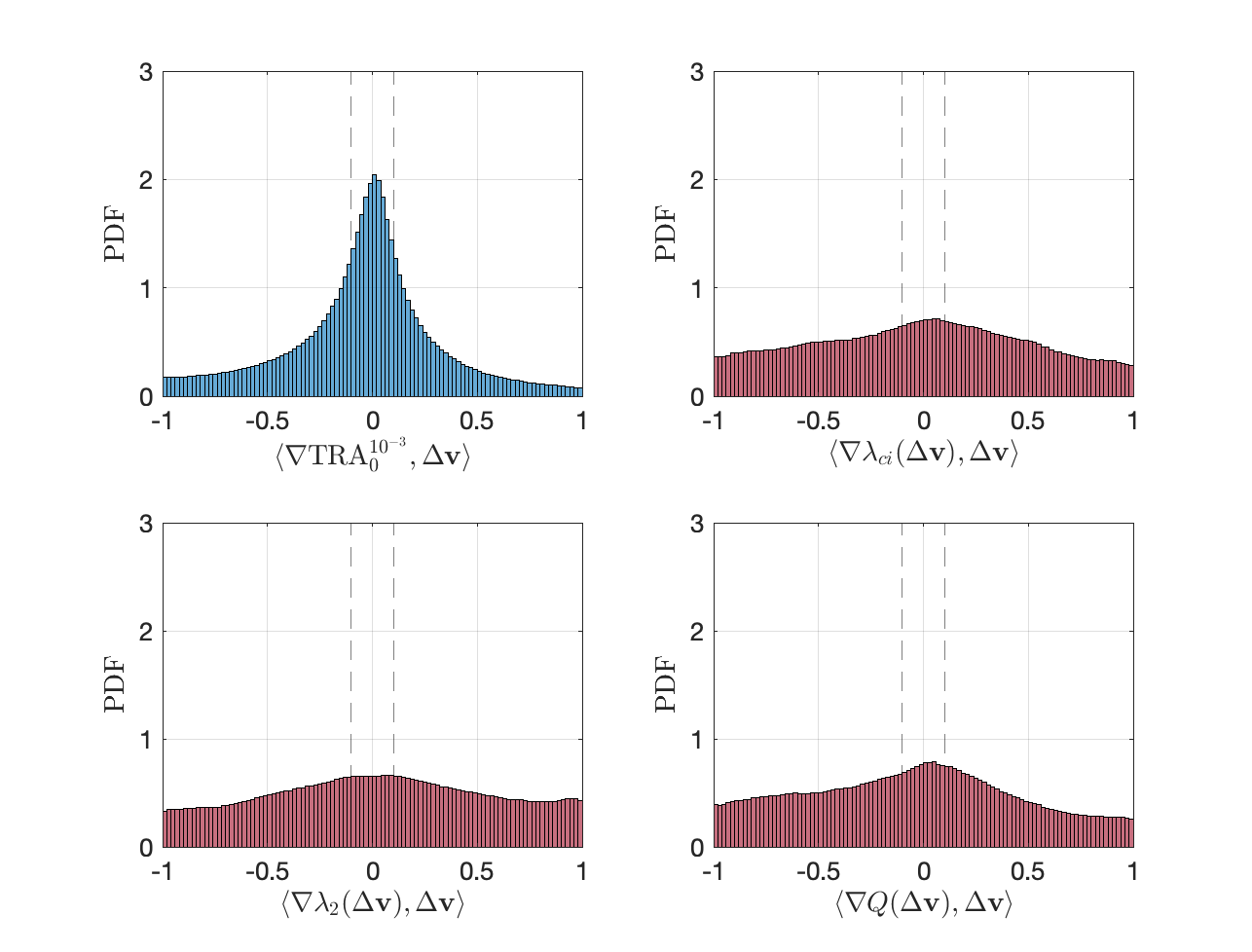}}
 \caption{Same as Fig. \ref{fig:InnerProd_Vel}, but with the $\lambda_{ci}$,
$\lambda_{2}$ and $Q$ metrics computed for the barrier equation
(\ref{eq:instant barrier}).}

\label{fig:InnerProd_Barrier}
\end{figure}

As the linear-momentum barrier vector field $\Delta\mathbf{v}$ is
objective, extracting the classic $\lambda_{ci}$, $\lambda_{2}$
and $Q$ level surfaces from the $\Delta\mathbf{v}$ field would also
be an objective procedure, unlike extracting these level surfaces
from the velocity field $\mathbf{v}$, as originally intended for
these diagnostics. To see if such an objectivization would benefit
these criteria, we have carried out the same statistical structure-tangency
analysis with the $\lambda_{ci}$, $\lambda_{2}$ and $Q$ metrics
now applied to the barrier field $\Delta\mathbf{v}$. While the
resulting PDFs in Figure \ref{fig:InnerProd_Barrier} now show a moderate
rise near zero for each classic vortex diagnostic, TRA level-sets
still outperform the other diagnostics in their ability to
block the viscous transport of momentum. This suggests that TRA level
sets are overall close approximations to the perfect momentum barriers
formed by invariant manifolds of the barrier equation (\ref{eq:instant barrier}).

\subsection{Momentum transport barrier interfaces}
\label{Section:MTB}

\begin{figure}
\centerline{\includegraphics[scale=0.175]{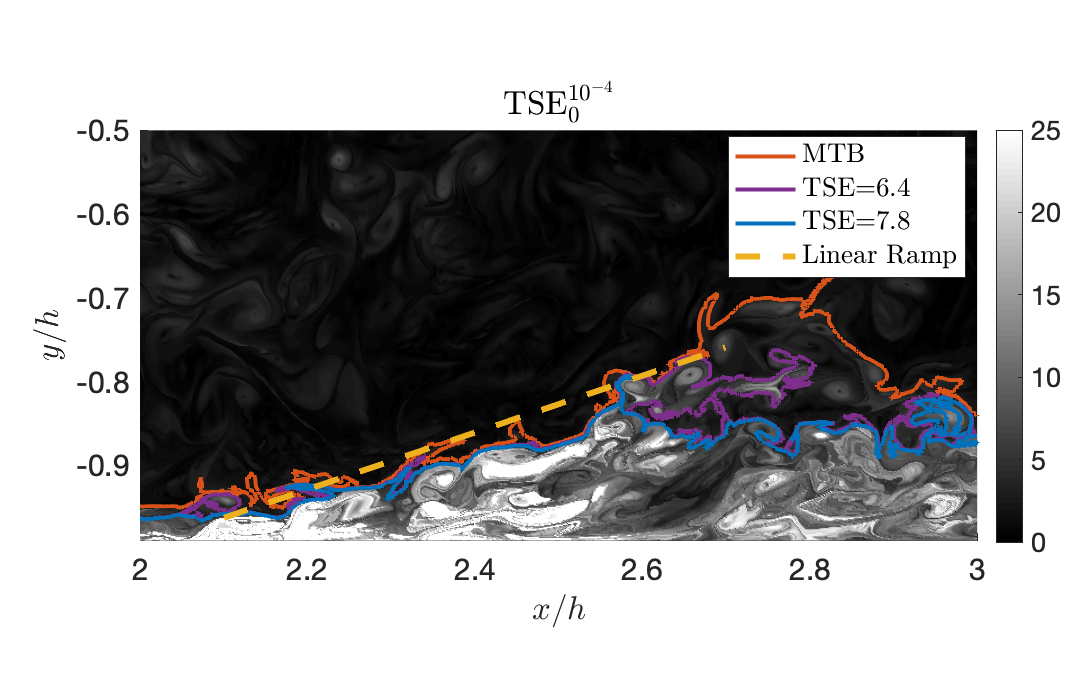}}
 \caption{TSE contours with the smoothest contour selected as an MTB. A linear ramp identified by a linear-fitting search is also drawn in. Note that the MTB is more effective at contouring around the outside of vortices whereas the other two TSE contours filament and penetrate the interiors of vortices.}
\label{fig:InterfaceChoice}
\end{figure}

We have, so far, used level surfaces of the TRA field for the approximate visualization
of vortices with a perfect instantaneous momentum-trapping property.
We can also use the TSE field to interrogate the momentum barrier equation more
globally in order to locate momentum-blocking interfaces between more
and less turbulent areas of the flow. This can be achieved by constructing
streamsurfaces of the barrier equation that partition the flow into wall-parallel domains with boundaries that span the channel in the wall-parallel directions. We use TSE fields to aid in this interface identification as their level surfaces separate regions of distinct degrees of stretching in a manner analogous to hyperbolic invariant manifolds. We use the single smoothest interface as this provides the barrier with the least folding. As shown in Figure \ref{fig:InterfaceChoice}, there are multiple spanning TSE interfaces in a given domain. These interfaces contour around vortices that compose large scale motions. After analyzing 100 temporal frames of TSE fields, we find that using the average straightest contours allows us to connect the outermost contour around these vortices. This selection can be modified for other flow-specific criteria, and a statistical analysis of all candidate TSE interfaces is conducted in Section \ref{sec:Statistics}.

We now describe a simple, automated algorithm
for identifying such objective momentum transport barriers (MTB) 
as a physics-based alternative to the currently used TNTI or UMZ interface
identification processes \citep{Adrian2000, DaSilva2014, DeSilva2015}. This algorithm is also sketched in Algorithm 2.
As with our vortex identification, when using the trajectory decorrelation time for TSE calculations, we only use one free parameter in our MTB algorithm: the choice of spatial resolution, which is approximately 1 viscous length in all directions. This ultimately controls the level of detail in the MTB interfaces and their momentum-blocking ability. In contrast to common TNTI or UMZ level-surface approaches, we can progressively improve our momentum-barrier identifications by calculating TSE at progressively finer resolutions beyond the underlying velocity grid. This improves surface triangulations and enhances tangency with invariant manifolds in the barrier field (\ref{eq:instant barrier}).

\begin{table}
\def~{\hphantom{0}}
\begin{tabularx}{\linewidth}{|l|X|}
{} & {\bf Algorithm 2: Extracting MTB} \\
1 & For a co-dimension 0 domain of interest $U\subset \mathbb{R}^3$, select a grid of initial conditions $\mathbf{x}_0\in U$. \\
  2& Determine the optimal barrier time $\epsilon_0$ using TSE calculations (instead of TRA) in Steps 2-4 in the vortex core extraction algorithm. \\
3 & For each streamwise-wall-normal hyperplane $V_z$ that intersects $U$, extract iso-contours in $V_z$ for the range of TSE values present in $V_z$. Record only the contours $\left\{ \gamma_{j}\right\}$ that span the full streamwise width of the domain. \\ 
4 & Refine $\left\{ \gamma_{j}\right\}$ to corresponding TSE values with contours that span the domain for every $V_z$. \\
5& Select the TSE value such that the corresponding $\gamma_j$ span each $V_z$ with minimal length. \\
6 & Extract the connected TSE isosurface corresponding to this TSE value that intersects the $\gamma_j$ of interest.
           \end{tabularx}
  \label{tab:kd}
\end{table}

We begin this algorithm by calculating the TSE for the active barrier
equation (\ref{eq:instant barrier}) on a 3D domain of interest. Here we use the decorrelation time as determined for the wall-proximal (outer) quarter of the channel ($|y|\ge0.75$) in an effort to visualize the most turbulent and complex momentum blocking structures ($s_N=10^{-4}$) emanating from the wall. Through a sensitivity analysis, we found MTB identification to be consistent over six orders of magnitude of $s_N$, but the visual clarity of near-wall structures begins to deteriorate away from our chosen $s_N$. Note that these six orders of magnitude relate to the integration time of a trajectory in the barrier field equations, and not to a physical spatial or temporal dimension. For example, a longer integration time may allow initially adjacent particles to separate to a larger degree, or for a trajectory in a vortex core to circulate more, thus allowing for more distinct TSE and TRA contours, but not necessarily a different size of spatial features. As well, the $s_N=0$ limit exists, providing an estimate of the structures detailed here. The decorrelation time for visualization close to the wall is shorter than the value $s_N=10^{-3}$ calculated for our focus on vortices near the center of the channel as the magnitude of the $\Delta\mathbf{v}$ is much greater near the wall and a shorter integration time is sufficient to obtain the same degree of detail.

\begin{figure}
\centerline{\includegraphics[scale=0.175]{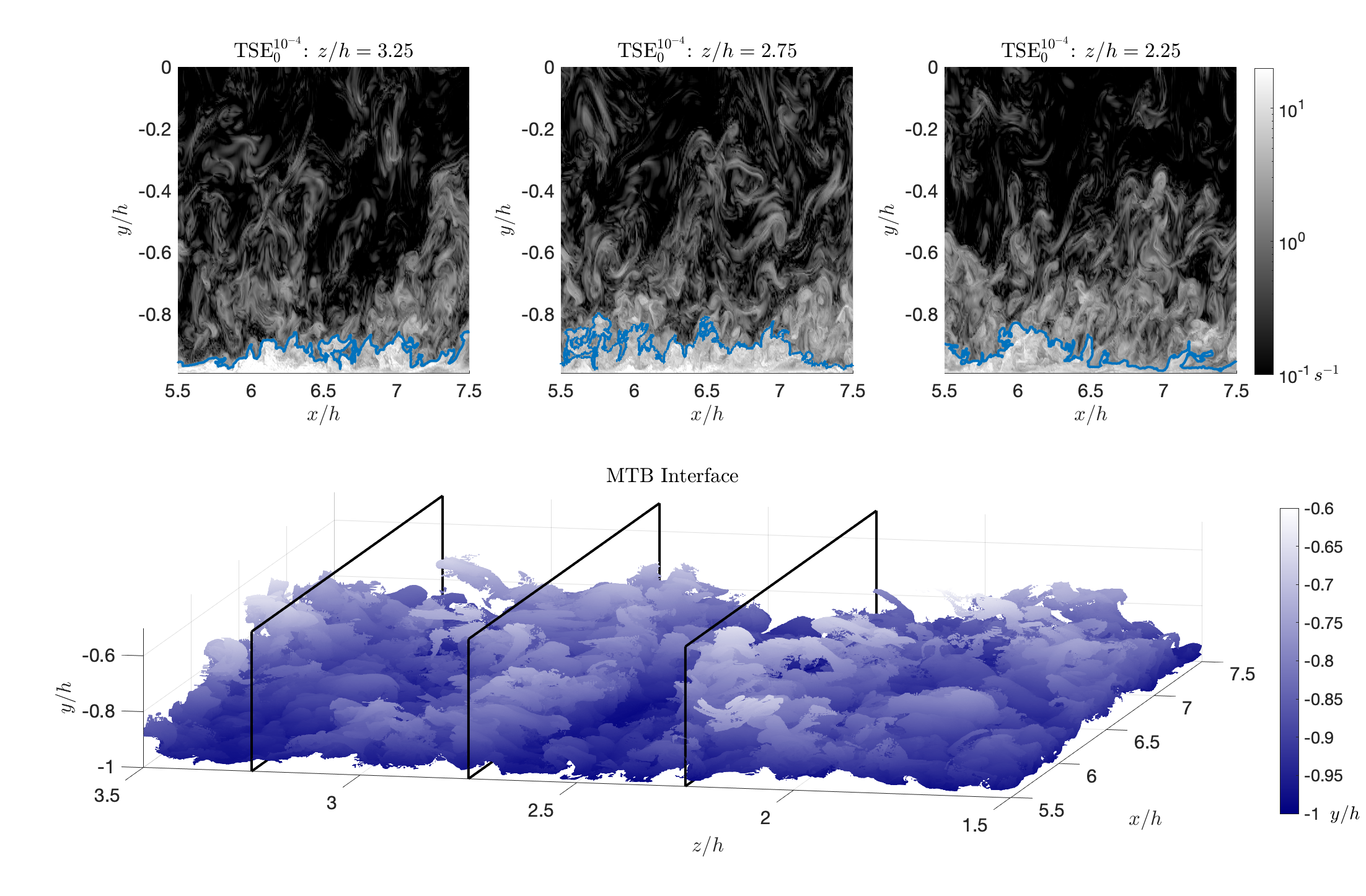}}
 \caption{Top: Process of selecting the TSE contour that spans a domain of interest
most efficiently among many adjacent 2D slices. Bottom: Corresponding TSE-derived momentum transport barrier (MTB) colored by y-values.}
\label{fig:TNTI_Process}
\end{figure}

To reduce the computational burden of finding the smoothest domain-spanning interface, we first perform a series of 2D approximations. We select a set of
$n\geq1$ streamwise-wall-normal planes and identify the
shortest TSE contour in each plane that divides the plane into a lower
(wall-adjacent) and upper (central) region. For example, see the candidate spanning contours at $\mathrm{TSE}=3.11$ for multiple $z$ planes in the top row of Figure \ref{fig:TNTI_Process}. Each such shortest, in-plane contour has a corresponding TSE
value whose corresponding 2D TSE level surface is a candidate
for an MTB interface. Of these candidate surfaces, we finally select
the 2D TSE level surface whose intersection curves with the $n$ streamwise-wall-normal
planes have the lowest maximum length. This procedure, therefore,
yields the MTB interface as the TSE level surface that divides the
flow roughly into parallel near-wall and mean-flow regions while maintaining
as low a curvature as possible. The corresponding lower half-channel MTB determined from this algorithm can be see in the lower plot of Figure \ref{fig:TNTI_Process}. The boundaries of the three upper planes of contour investigation are drawn in black. More involved algorithms targeting
the same objective can certainly be devised but will likely come with
increased computational cost. 

The MTB interface in Figure \ref{fig:TNTI_Process} is a complex structure
that reveals connections of multiple vortices as they collect and migrate
from the channel wall to the less turbulent center of the channel. The MTB has been shaded by the distance from the lower channel wall to help illuminate heterogeneity in the wall-normal extent of this MTB. Qualitatively familiar material features from smoke and dye experiments can be seen in this objective barrier. The organization of characteristic interface eddies documented by \citet{Falco1977} and \citet{Head1981} are evident along the MTB as well as the large scale streamwise streaky structures of \citet{Kline1967}, with a spanwise spacing of approximately $0.75h$. There is also a striking similarity to scalar concentrations visualized in jet-driven TNTI \citep[e.g.][]{Westerweel2009}

The MTB interface obtained in this fashion approximates the flattest
invariant manifold of the barrier equation that divides the flow into
two disjoint quasi-wall-parallel layers with minimal diffusive momentum
transport between them. The low-curvature requirement in the construction
of this interface forces it to avoid highly turbulent regions and
effectively connect outer regions of the momentum-trapping vortices
that we have already discussed. 

\begin{figure}
\centerline{\includegraphics[scale=0.25]{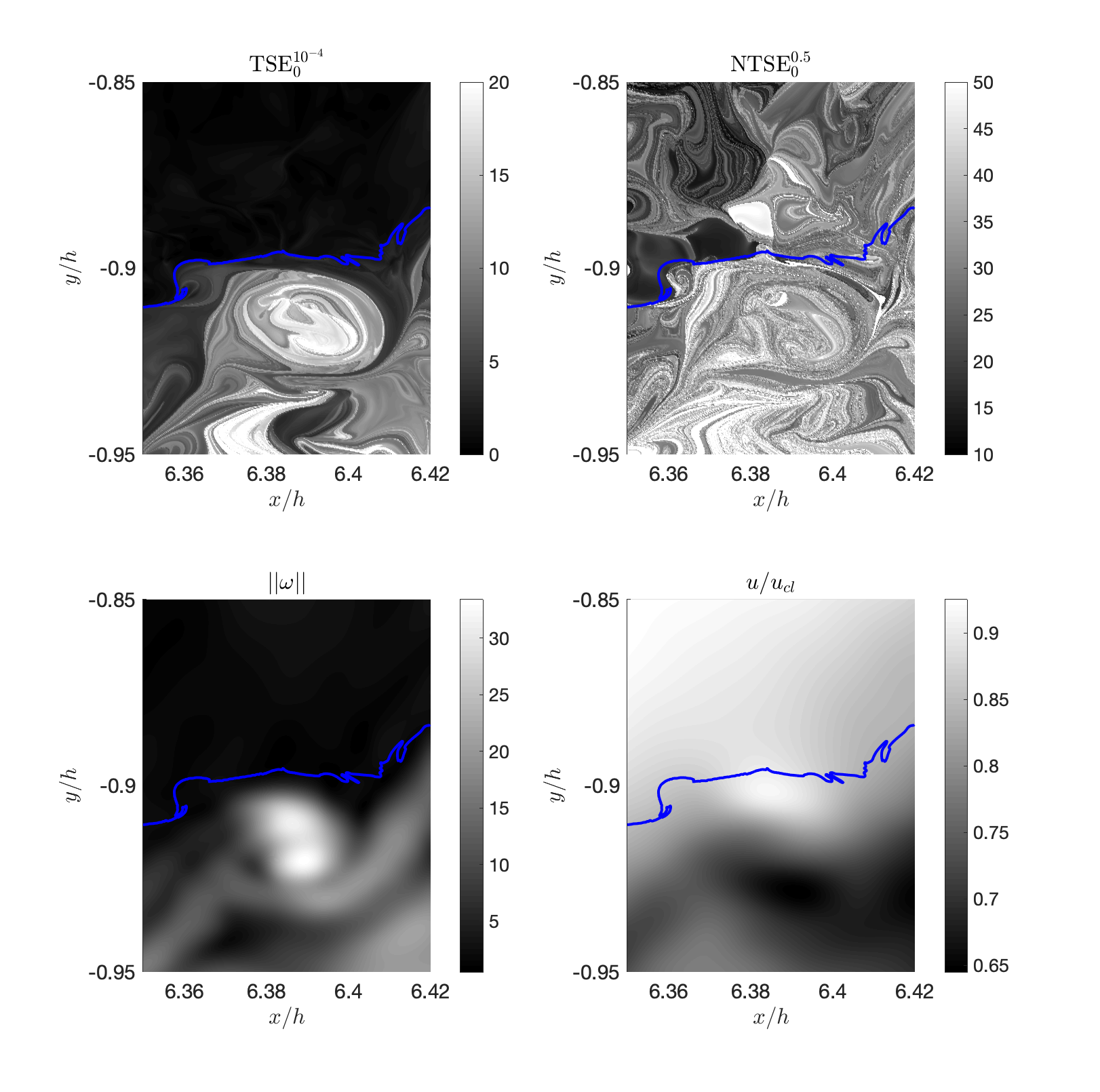}}
 \caption{CW from the top left: The TSE field, the NTSE field for the unit barrier field, the streamwise velocity
$u$ normalized by the centerline velocity $u_{cl}$, and the vorticity norm
$\left\Vert \boldsymbol{\omega}\right\Vert $. The MTB interface is superimposed in all plots
as a blue curve. All image data was computed and displayed at the
same spatial resolution. }
\label{fig:TNTI1}
\end{figure}

To illustrate this behavior, in Figure \ref{fig:TNTI1} we focus on a small subdomain of Figure \ref{fig:TNTI_Process} and compare our interface with two commonly used UMZ and TNTI identification diagnostics, vorticity and normalized streamwise velocity. While high-shear regions have been associated with UMZ interfaces \citep{Hunt1999, Adrian2000}, and have been used to help separate UMZs \citep{Eisma2015}, high-shear zones do not typically span the entire domain or provide a complete zonal separation of the flow. We thus conduct our comparison of domain-separating MTBs with other level-sets that span the domain. We also include in this comparison the NTSE field.

 The top-left panel in Figure \ref{fig:TNTI1} shows TSE values calculated
for initial conditions on the $z=2.75h$ plane with the interface
drawn in blue. The interface effectively contours around the outside of a strongly stretching spiral feature and separates the flow into a less turbulent and more turbulent region. The NTSE field in the top-right panel reveals many of the same structures tangent to the MTB with additional weaker features aligned above the interface. Slight differences between contours in TSE and NTSE fields on the edge of the domain can be attributed to fixed time step integration effects.

The bottom-right panel shows the streamwise velocity, a scalar field
whose contours are broadly used for extraction of uniform momentum
zones in wall turbulence \citep[e.g.][]{Adrian2000, Kwon2014, DeSilva2015}.
Following the procedure of \citet{Kwon2014}, we have normalized the
velocities by the channel centerline velocity. We note that the predominant
gradient in the velocity field follows approximately through the center of the vortical feature, and the velocity contours are parallel with the MTB in only
a small region of the domain. We will further explore the differences
between UMZ interfaces MTB interfaces below.

Lastly, the bottom-left panel of Figure \ref{fig:TNTI1} shows the
vorticity norm in the same domain, a diagnostic commonly used in TNTI
identification \citep[see][]{Holzner2006, DaSilva2014}. This plot provides
a similar but simpler picture of the flow dynamics in comparison with
TSE, but the details do not indicate exactly where an interface should
be drawn. There is no localized shear-driven vorticity peak along the turbulent interface, as is sometimes possible at the TNTI
\citep{DaSilva2014}. As well, the change in vorticity is quite gradual, not giving a clear jump in the vorticity PDF
as is often used to separate rotational and irrotational flow
fields. For these reasons, vorticity is not commonly used for internal
interface diagnostics, but is included here as it does provide a closer
comparison with momentum barrier field behavior than the streamwise
velocity (UMZ) visualization. As well, much of UMZ theory relies on
an analogy with jumps across turbulent/non-turbulent interfaces.

\begin{figure}
\centerline{\includegraphics[scale=0.2]{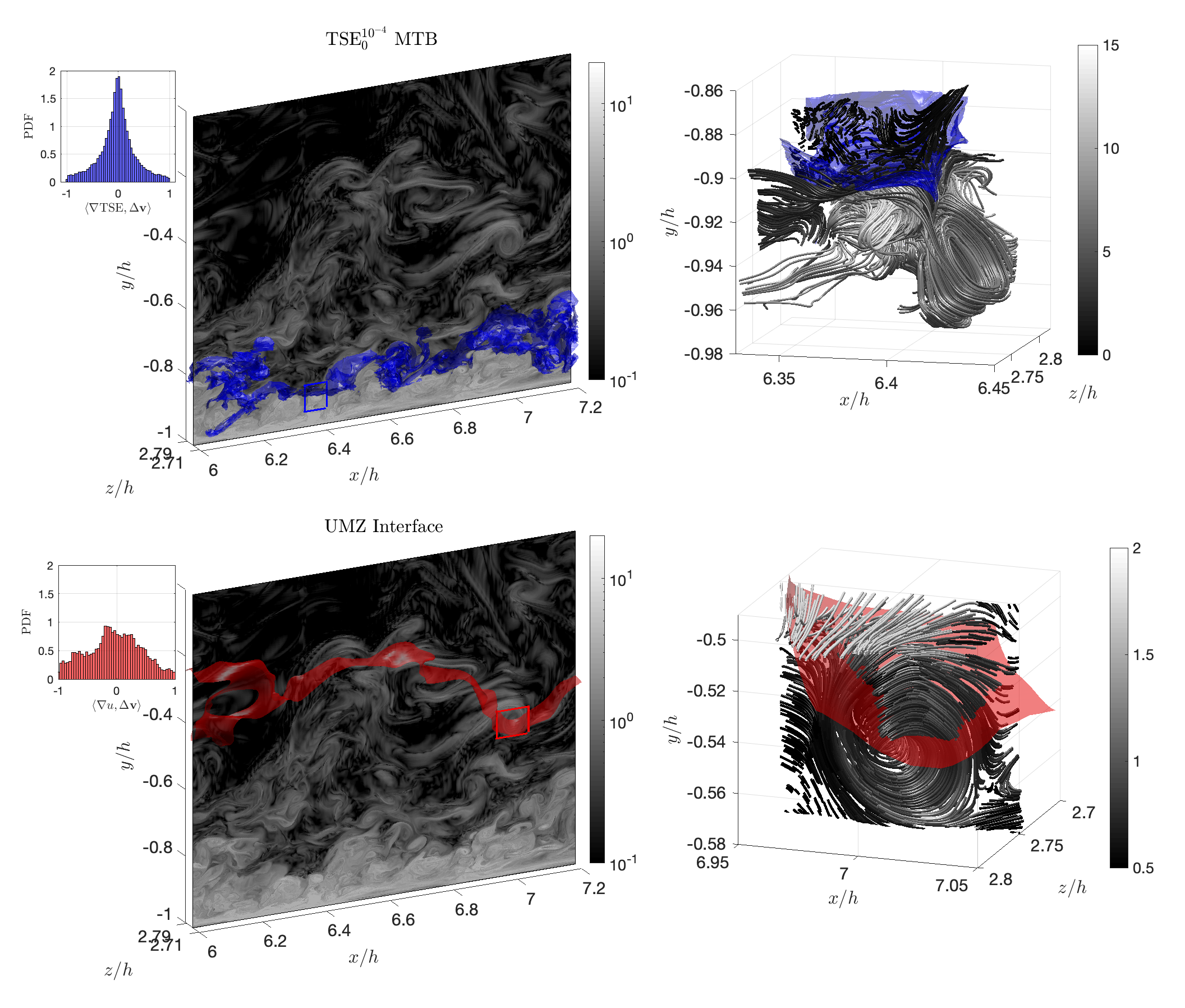}}
 \caption{Top Left: TSE field and MTB interface. Bottom Left: TSE field and UMZ interface. Right: Corresponding interfaces and linear momentum barrier field trajectories colored by TSE. Inset in each plot are the corresponding PDFs of surface normal innerproducts with the barrier field indicating improved momentum transport limiting abilities of the MTB. }
\label{fig:StraightestContour_Downsample_Compare}
\end{figure}

The ability of the TSE field to identify divergent behavior of barrier streamlines and appropriately locate the MTB is illustrated in Figure \ref{fig:StraightestContour_Downsample_Compare}. Here, we expand our focus from Figure \ref{fig:TNTI1} to a narrow subsection of the channel. The left column shows the local geometry of the
MTB (blue) and UMZ (red) interfaces near the $z=2.75$ plane. The right column shows a zoomed in view of the interfaces in the domain of Figure \ref{fig:TNTI1} and the nearby
trajectories of the barrier equation (\ref{eq:instant barrier}).

Trajectories are shaded by their TSE values. For the MTB interface, there is an obvious separation of minimally-stretching black trajectories in a less turbulent region above the blue interface and the multiple spiraling grey and white vortices below the interface. The MTB is also tangent to the barrier field streamlines, indicating the correct orientation with respect to the momentum barrier streamsurfaces. This is further confirmed in the inner-product probability distribution inset for the MTB on the left. Similar to the findings for TRA level-surfaces around vortices in Figure \ref{fig:InnerProd_Vel}, the geometry of TSE level-surface interfaces is largely tangent to perfectly computed zero-flux linear momentum barriers.
 
The bottom-left panel of Figure \ref{fig:StraightestContour_Downsample_Compare} shows the universal
$u/u_{cl}=0.95$ ``quiescent core" UMZ interface calculated over the $1.2h$ streamwise extent, as suggested by \citet{Kwon2014}.
The MTB and UMZ interfaces differ greatly, but the UMZ interface does hint at being influenced by the TSE contours on the right side of the domain. This can be see in the inner-product PDF on the left with one such highlighted example shown in the bottom-right panel. The tangency PDF for the UMZ shows a slight peak around zero corresponding to near parallel points on the right side of the domain. Barrier field streamlines around a vortex feature, however, are seen to be perpendicular to the UMZ interface, thus maximizing transport. This is in direct contrast with
our automated MTB interface algorithm that has clearly succeeded in approximating
a repelling (and hence structurally stable) invariant manifold of
the incompressible barrier equation. Of note, the UMZ interface transects the vortex core, as also seen by velocity contours in Figure \ref{fig:TNTI1}. The UMZ interface is clearly not an organizing structure that blocks momentum transport, but these TSE fields and MTB may begin to give insight into the physical origins of the commonly seen statistical signatures that have supported the ubiquitous use of velocity isosurfaces.

While TSE level surfaces that span the domain at the height of the UMZ do exist, the nearby TSE structures contour and fold around many hyperbolic and elliptic barriers (such as those seen in Figure (\ref{fig:ChannelVis})) and do not provide a clear interface between distinct wall-parallel flow zones. We suspect that, in the neighborhood of the quiescent core interface, a simple quasi-planar structure that blocks wall-transverse momentum flux may not actually exist as the elliptic and hyperbolic barrier field manifolds are less densely concentrated. The transverse intersections of velocity level surfaces with invariant manifolds of (\ref{eq:instant barrier}) in the PDF of Figure \ref{fig:StraightestContour_Downsample_Compare} suggest any possible correlations between velocity level surfaces and true momentum barriers are insignificant for predictive ability.

\begin{figure}
\centerline{\includegraphics[scale=0.17]{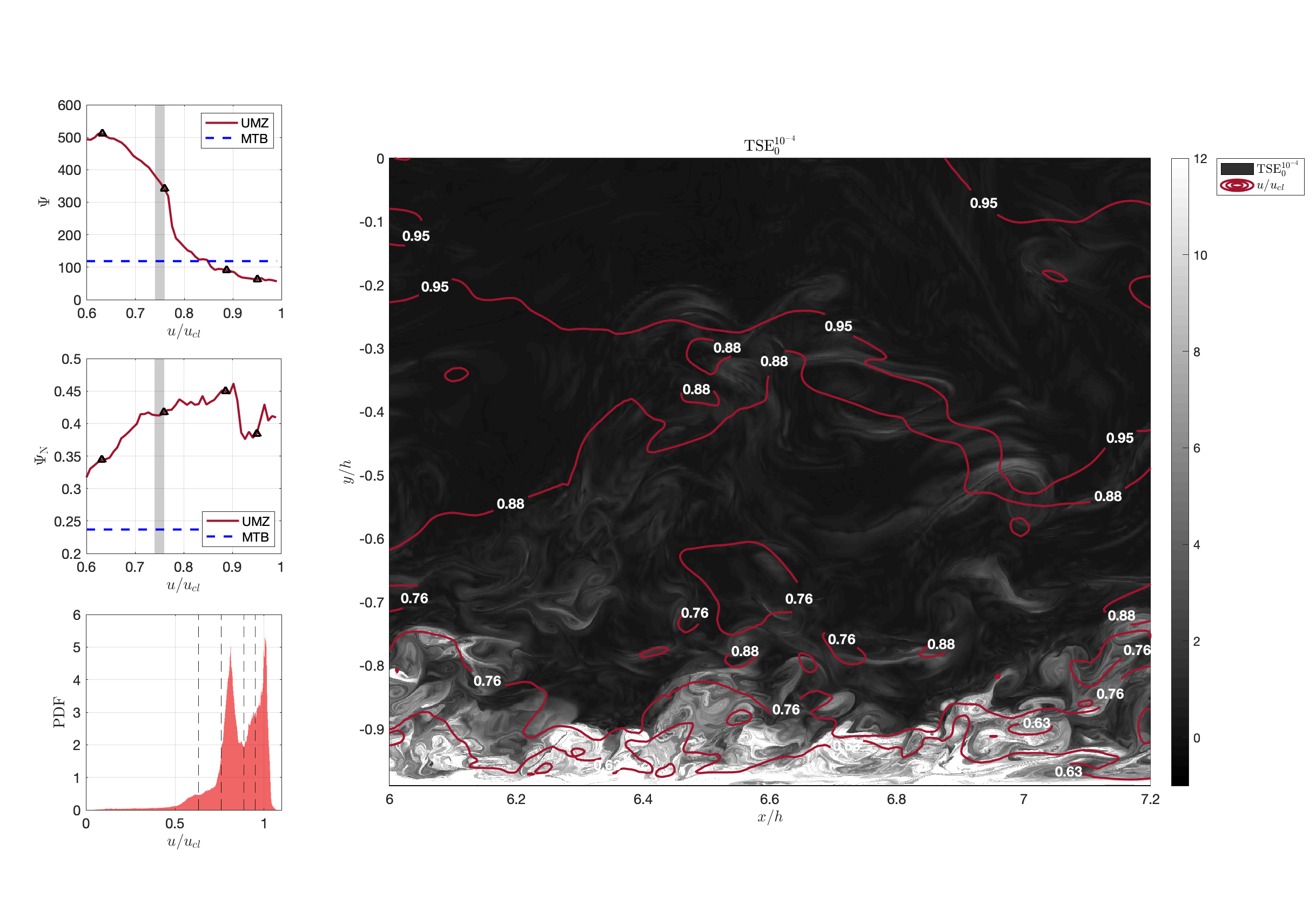}}
 \caption{Top left: The surface-area normalized instantaneous viscous momentum flux $\Psi$
through streamwise velocity isosurfaces (UMZ interfaces). Middle left: $\Psi$ in unit barrier field
for streamwise velocity isosurfaces. Bottom left: PDF of $u/u_{cl}$. Right: $u/u_{cl}$
contours (with their values indicated) superimposed in red over
the TSE field in the $z=2.75$ plane.}
\label{fig:TNTI3}
\end{figure}

To assess this relationship more broadly, we compare the momentum transport through a range of UMZ interfaces and the
MTB in Figure \ref{fig:TNTI3}. The right panel shows the same TSE field for the $z=2.75h$ plane as in Figure \ref{fig:TNTI1} with overlaid streamwise velocity contours. As there are numerous approaches to extracting the best high-shear streamwise velocity isosurface for UMZ identification \citep[see, e.g.][]{DeSilva2015,Fan2019}, we directly compare our MTB with a range of simply connected velocity level surfaces from the near-wall region well into the channel core. The top left panel shows the geometric diffusive flux across this range of interfaces in red, and our MTB flux in blue. The gray shading indicates the $u/u_{cl}$ isosurfaces closest to the MTB in the domain. As with the non-objective diagnostics in Section \ref{Section:Vortices}, these velocity isosurfaces exhibit around four times the flux as our nearby objective MTB.

As we increase the velocity of the UMZ interface, we also move away from the wall and into a region with smaller $\Delta \mathbf{v}$ vectors. While we have seen that the quiescent core UMZ interface is largely transverse to momentum transport barriers, its diffusive momentum flux is smaller than for our MTB.  We believe this is due to its location in a region with less turbulence and less flux, and not an ability to limit momentum transport. We thus calculate the normalized geometric flux (tangency measure) $\Psi_N$ from (\ref{eq:norm geometric flux}). In the middle panel, we present $\Psi_N$ for the same UMZ candidates. This shows the clear advantage of identifying internal momentum blocking interfaces with the MTB approach over any streamwise velocity isosurface. Triangles in the two flux plots mark several velocity isosurfaces of interest: the highest flux (0.63), the UMZ closest to the MTB (0.76), the trough between the two PDF peaks (0.88), and the quiescent core UMZ of \citet{Kwon2014} (0.95). The intersection of these four $u/u_{cl}$ isosurfaces with the $z=2.75h$ plane are drawn in the right
panel.

UMZ theory suggests that momentum is organized into zonal-like structures inside of which there is relatively uniform momentum. These zones are separated by thin viscous shear layers that are often thought to concentrate spanwise vorticity \citep{Meinhart1995}. Transport of momentum is concentrated to the viscous-inertial layers which separate the zones with the interfaces between them identifiable
from streamwise velocity histograms. Whether in boundary layers or channel flows, UMZ interfaces
have been defined as minima between peaks in the PDF for $u$, and correlated with high-shear.
The bottom left panel of Figure \ref{fig:TNTI3} shows a PDF for the
bottom half of the channel flow over the 3D domain of focus. These
PDFs are known to be highly sensitive to the size and location of the domain of
investigation but there is no consensus available on their use in 2D
or 3D analysis \citep{Fan2019}. The level set values contoured in
the right panel are marked on the PDF with dashed lines.

It can be seen in Figure \ref{fig:TNTI3} that streamwise velocity contours interfaces with relatively
low diffusive momentum flux do not correspond with momentum transport limiting structures. Indeed, all $u/u_{cl}$ interfaces
 travel transverse to multiple vortex cores (including that highlighted in Figure \ref{fig:StraightestContour_Downsample_Compare}) and meander in and
out of high stretching regions.  In the $\Psi$ and $\Psi_N$ plots, we can see the surface derived from the minima between the two largest PDF peaks, $u/u_{cl}=0.88$, does not provide an even locally minimal momentum flux as has been suggested. The steady increase in $\Psi_{\mathrm{N}}$ and decrease in $\Psi$ for increasing $u$ shows the dominant influence of barrier field magnitude on $\Psi$ and a lack of momentum barrier tangency. In fact, the spanning velocity level surface with minimum momentum flux occurs at $u/u_{cl}=0.99$ which is well inside the less turbulent region of the flow. Of note, the quiescent core, $u/u_{cl} =0.95$, interface does indeed appear adjacent to the dominant channel core PDF peak, as suggested
by \citet{Kwon2014}. However, the $\Psi$ and $\Psi_N$ and plots shows no indication of significant momentum transport organizing structures and no indication which velocity isosurfaces exist inside or on the border of a momentum-organizing zonal-structure. That is, there is no evidence of shear-layers of localized vorticity that concentrate or minimize momentum transport between UMZs.
 
\begin{figure}
\centerline{\includegraphics[scale=0.375]{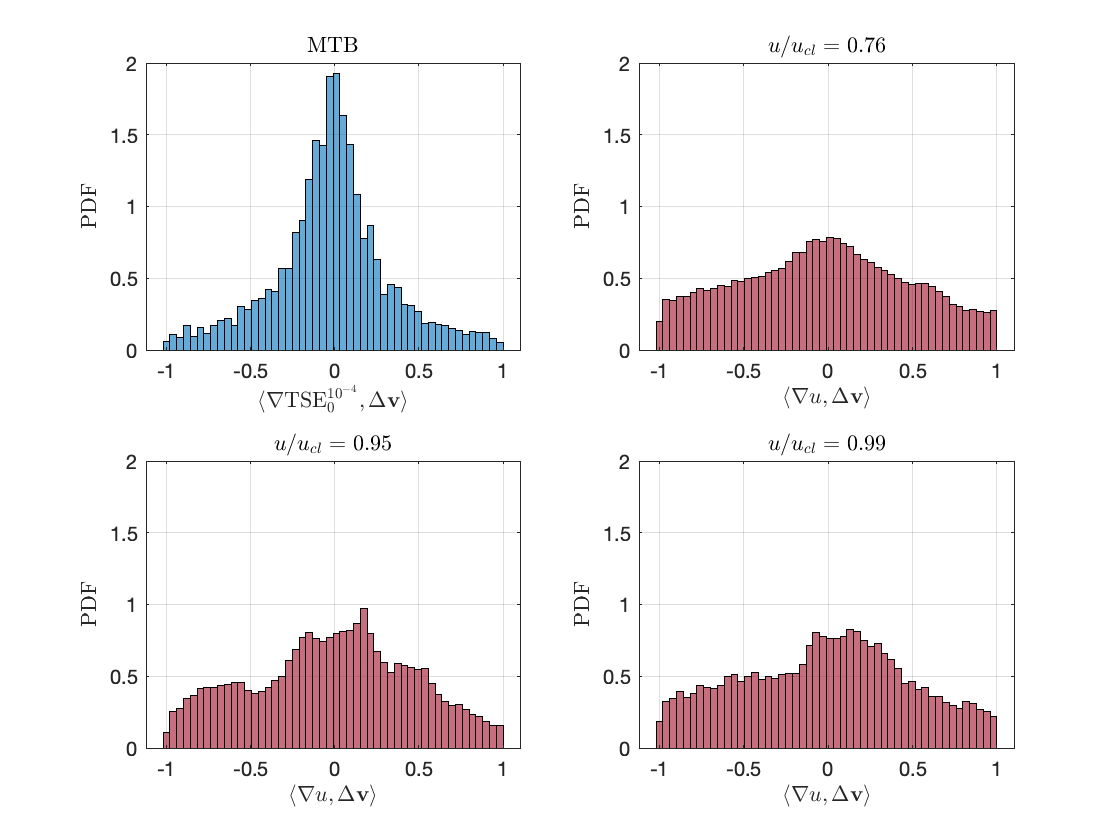}}
 \caption{Probability distributions of the normalized inner product of the momentum
barrier field vector, $\Delta\mathbf{v}$, and the isosurface normals
for an MTB and UMZ interfaces. There remains a clear singular peak around 0 in the TSE
PDF, indicating strong agreement between TSE surfaces and the underlying
momentum transport blocking interfaces both for vortices and for the
connecting regions between them. In contrast, there is significantly
more momentum transport across UMZ interfaces.}
\label{fig:TRA_UMZ_Angles}
\end{figure}

 Each panel in
Figure \ref{fig:TRA_UMZ_Angles} shows the normalized inner product of respective isosurface
normals with the momentum barrier field vector $\Delta\mathbf{v}$. 
As before, values around zero in the PDFs indicate an alignment of
isosurfaces with true streamsurfaces of the momentum barrier equation.
The top left panel shows a familiar peak around 0 for the MTB interface,
confirming that the low momentum flux calculated across this interface
is aided by its close alignment with a set of invariant manifolds of the barrier
equation. There are impressively few isonormal vectors lying outside
this peak even with the algorithm
balancing strict coincidence invariance under the barrier equation
with computational simplicity and moderate curvature.

In comparison, the PDF associated with the closest candidate UMZ interface to the MTB, $u/u_{cl}=0.76$,
 exhibits a subtle but much less pronounced peak around zero. Note
that this surfaces experiences more than three times the momentum flux across
it than the MTB interface. The other two velocity isosurfaces, including the quiescent cored interface, exhibit similar minor peaks around zero, with much more uniform
and random alignment of surface normals when compared to the momentum
barrier field vectors. This confirms our previous suspicion that flux is primarily
minimized across this surface because of its location in a less turbulent
region of the flow, and not because it behaves as a physically significant
barrier. This is somewhat unsurprising as a simple wall-parallel plane in the neighborhood of the $u/u_{cl}=0.95$ level surface has a lower geometric objective momentum flux than the quiescent core boundary. However, for each contour $u/u_{cl}=0.76,$ 0.88, and 0.95, there are small sections of near tangency with the underlying momentum barriers. While velocity contours and momentum barriers are clearly not equivalent, this again suggests that method of momentum barrier identification is likely a beneficial avenue for investigating statistical signatures.

We have found the diffusive momentum transport blocking abilities of MTBs in the momentum barrier fields to be consistent in time frames spanning the entire DNS simulations. Focusing on $1.2h \times 0.5h \times 0.2h$ domains, we calculated TSE fields for 25 frames, generating more than 100 GB of TSE data. The PDF of normalized flux angles consistently follows similar behavior to that seen Figure \ref{fig:TNTI3}, with an effective approximation to underlying invariant manifolds of the barrier equations. 

\subsection{Tracking features and temporal statistics} 
\label{sec:Statistics}

\begin{figure}
\centerline{\includegraphics[scale=0.375]{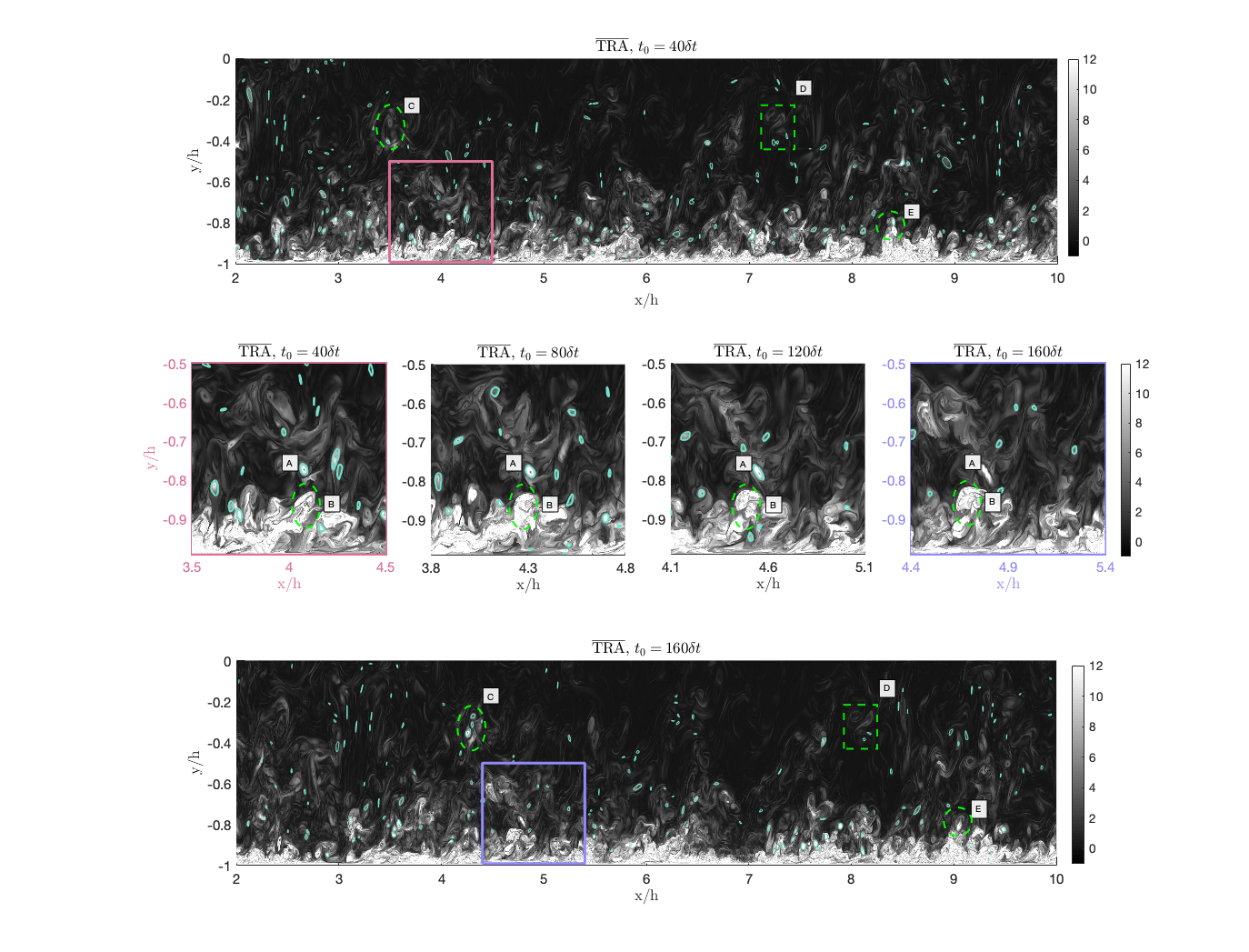}}
 \caption{Subsequent $\mathrm{\overline{TRA}}$ fields in the x-z plane spanning 120 channel flow-through times. Multiple features that can be tracked throughout the frames have been identified, with a zoom section and intermediate frames shown in the middle plots. Vortex boundaries identified as the outermost closed $\mathrm{\overline{TRA}}$ contours are also drawn in, many of which can be tracked between frames. }
\label{fig:Track_Features}
\end{figure}

The JHTDB Re=1000 channel flow spans 4,000 flow-through times ($\delta t$). We evaluated the frequency and recurrence of momentum-blocking vortex cores and MTBs in 100 frames with temporal spacing of $40\delta t$ spanning the entire available simulation time. Each frame is adjacent to the lower channel boundary, spanning $h \times 8h$ in the streamwise-wall-normal plane. This required analyzing 800 GB of velocity data, and resulted in 30 GB of TRA and TSE fields, which are available in the supplementary material, as is an animation of the time evolution of TSE frames.

A visual inspection of subsequent TRA and TSE fields reveals that many easily identifiable Eulerian features remain coherent during fluid advection and can be tracked from one frame to the next. One such example of this feature tracking is shown for the $z=3h$ plane in Figure \ref{fig:Track_Features}. The top and bottom plots show TRA fields at $t_0=40\delta t$ and $t_0=160\delta t$, with the middle four frames tracking features in zoomed areas for intermediate timesteps. Some features of interest are labeled A-E; individual vortex cores following Algorithm 1 are highlighted in turquoise. 

The vortex immediately to the right of the label A in the intermediate frames can be followed from $40\delta t$ to $160\delta t$, though the vortex eventually moves out of frame so that a circular contour is no longer extracted at $160\delta t$. The region labeled B is a cluster of tangled vortices near the top of a wall-connected bulge that is moving away from the wall. Because of the detail possible with TRA, individual components of B can be distinguished and followed between adjacent frames. From $40\delta t$ to $160\delta t$, vortices A and B remain identifiable while being advected nearly a full channel half-width down the channel.

In the wide-views at the top and bottom of the Figure, three additional features are identified and tracked. There are two vortex clusters, C and D, sitting approximately $0.7h$ from the wall in a much faster moving part of the channel. Vortex E is also tracked as the uppermost vortex on a bulge that has migrated approximately $0.75h$ down the channel. Note that the advection velocity of these five features is fairly consistent though their wall-normal locations vary by $0.6h$. The visual matching of these features was verified by advecting fluid particles in the DNS velocity field over this time window. While there was some minimal deformation of vortices over the period (not displayed), the advected structures and latter Eulerian detections indeed matched. This is promising for not only Eulerian studies of boundary layer structure evolution in great detail, but also the possible use of material momentum barriers from the Lagrangian theory of \citet{Haller2020}. A video animation of the TRA fields in this plane spanning the entire DNS database be found in the supplementary movie.

\begin{figure}
\centerline{\includegraphics[scale=0.22]{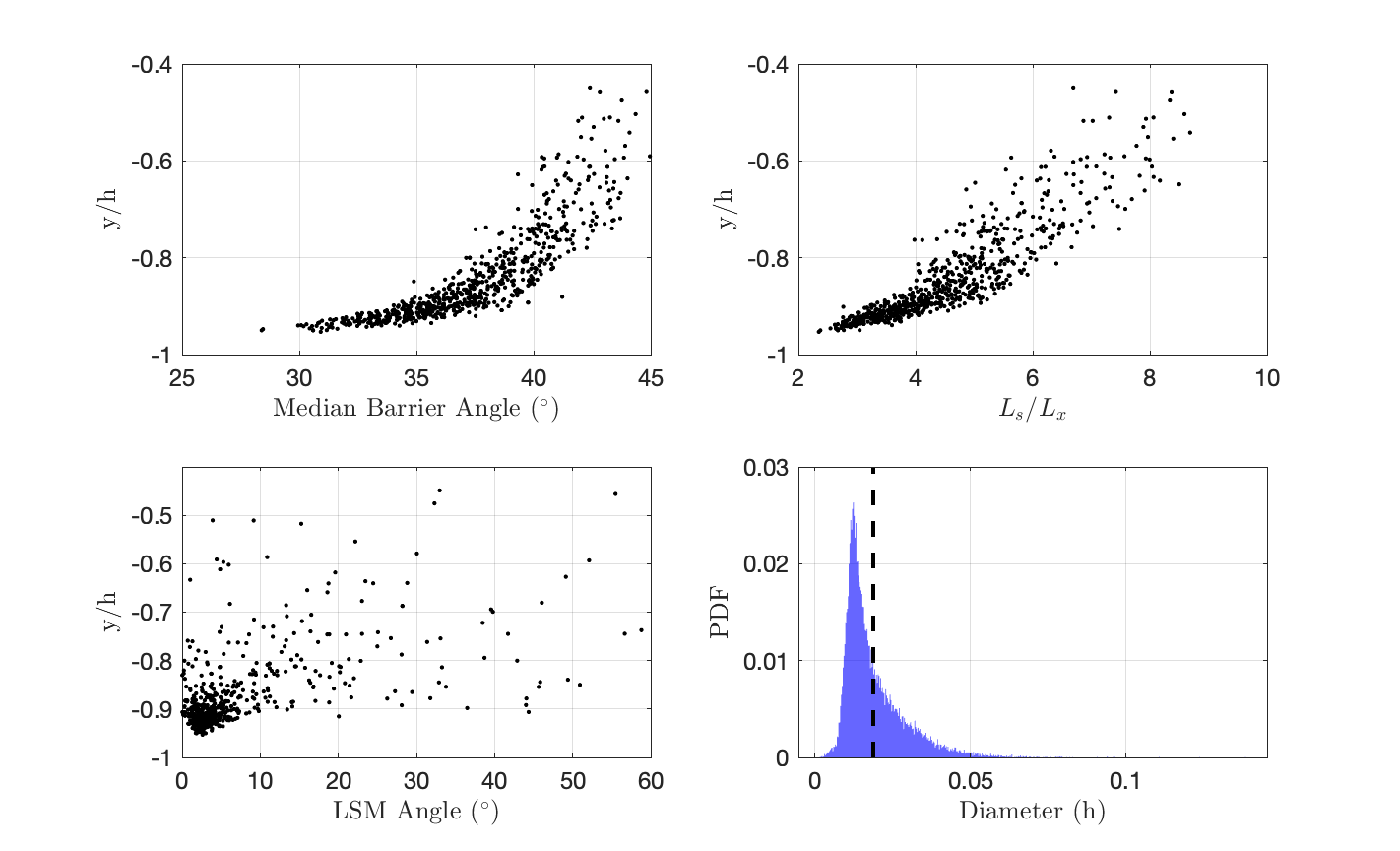}}
 \caption{Interface angles and radii of vortex cores (heads) over in 3D frames spanning 4,000 flow-through times.}
\label{fig:Angles}
\end{figure}

By statistically analyzing the entire temporal extent of the channel flow simulation, we found similar behavior and characteristics for vortices and MTBs. We analyzed all TSE contours that spanned the $8h$ domain and found at least one such contour existed in each frame. The total number identified depends on the choice of the number of contour values to extract. At this Reynolds number, these contours were typically closely adjacent, as in Figure \ref{fig:InterfaceChoice}, though more distinct layering is likely at higher $Re$ as noted by \citet{DeSilva2017}. We include all TSE contours in our statistical analysis as candidate MTBs in light of potential advances in selection criteria and find this does not skew our statistics or introduce outliers. The median positive angle of each MTB was calculated and is displayed with respect to the median height of that momentum interface in the top-left panel of Figure \ref{fig:Angles}. TSE contours follow the boundary of individual vortices and provide a measure of the incline angle of individual hairpins or typical eddies as a function of height. We find growth away from the wall in a similar fashion to existing attached eddy (hairpin vortex packet) models with angles ranging from $30^\circ$ up to $45^\circ$ further from the wall \citep{Marusic2001}.

Next, we measured the unit length per streamwise distance ($L_s/L_x$) of MTBs, similar to \citet{Chauhan2014} and \citet{DeSilva2017}. In the top-right plot of Figure \ref{fig:Angles}, there is a clear increasing trend for barriers further from the wall. This agrees with the assessment in previous studies that, at greater distances from the wall, larger scale bulges and valleys contribute to greater $L_s/L_x$ \citep{Townsend1976, Perry1982, Adrian2000, DeSilva2017}. $L_s/L_x$ for our MTB exceeds the range of \citet{DeSilva2017} ($<3$), but our mean, $\left<L_s/L_x\right>=5.4$, is much closer to that found by \citet{Chauhan2014} for the TNTI. This is likely attributed both to physical differences as well as the greater detail and meandering stream-surfaces found in $\Delta\mathbf{v}$ when compared to the underlying velocity grid (e.g. Figure \ref{fig:ChannelVis}).

By identifying the outermost closed and convex TRA contours, we found momentum vortex with have a mean diameter of $0.019h$. The PDF of this diameter distribution is shown in the bottom-right panel of Figure \ref{fig:Angles} with the mean marked as a dashed line. These vortex cores are smaller in diameter than the observed tracer eddies of \citet{Falco1974, Falco1977}, and slightly smaller than the $0.03-0.05\delta$ vortex heads identified by \citet{Adrian2000}. There are several possible reasons for this, the most probable being that the core vortical structure is smaller than its visible influence on the material (and velocity) field. We are restricting to closed and convex contours and thus eliminate the filamentation seen in smoke experiments. 

Another commonly referenced statistic in the boundary layer literature is the inclination angle, $\alpha$, that the envelope of packets of vortices make with the wall, also known as the structure angle or ramp angle. This has been frequently measured by two-point correlations, with limited guidance for quantifying spatially-resolved structures beyond the Radon transform used by \citet{Hommema2003}. We opt to measure the inclination angle of 0.6$h$-streamwise-length linear segments that fit our contours of interest with $\mathrm{RMS}<0.02$, the average diameter of our vortices. This length was chosen as it is less than the lower bound of the streamwise length of bulges reported in the literature, and approximately one half the streamwise extent necessary for converged channel flow UMZ statistics \citep{Kwon2014}. This provides sufficient streamwise range to find a good linear fit to a ramp without fitting to the small individual sub-components or the large scale motions. An example of a successful linear fitting is shown in Figure \ref{fig:InterfaceChoice}.

The average $\alpha$ value is found to be $8^\circ$ with the average ramp center approximately $0.18h$ away from the wall. There were approximately $75\%$ as many ramps identified as total number of MTBs. While these angles and heights are similar to the observations of \citet{Head1981}, and seminal UMZ work \citep[see, e.g.][]{Adrian2000}, there is significant scatter around these values as can be seen in the bottom-left of Figure \ref{fig:Angles}. This suggest the linear fit method should be refined further in future momentum barrier investigations.

\section{Conclusions}

Using the objective notion of diffusive momentum flux through a material
surface, we have developed an algorithm for locating frame-invariant
instantaneous barriers to linear momentum transport in near-wall turbulence.
This algorithm builds on the recent theory of active transport barriers
\citep{Haller2020}, which identifies momentum transport barriers
as structurally stable invariant manifolds (stream surfaces) of the
incompressible steady barrier equation (\ref{eq:instant barrier}).
Our algorithms extract these invariant manifolds directly and approximate them by level surfaces of recently developed coherent structure diagnostics,
the trajectory rotation average (TRA) and trajectory stretching exponent (TSE) proposed by \cite{Haller2021}.
Other objective LCS diagnostics can also be used to identify invariant
manifolds of (\ref{eq:instant barrier}), but TRA and TSE were used here due
to their computational simplicity as single-trajectory-based diagnostic
fields and their ability to designate clean boundaries. We also introduce the normalized NTRA and NTSE, which provide unprecedented detail for coherent structures, with equal fidelity in highly turbulent and less turbulent zones.

Our procedure targets both momentum-trapping vortices in the boundary
layer and internal momentum-flux minimizing interfaces that locally define
the flow into near-wall and far-from-wall layers. Specifically, vortex
boundaries are identified as the outermost cylindrical level surfaces
of the TRA field. This procedure is free from the empirical threshold
values employed by classic, velocity-gradient-based vortex criteria
which provide observer-dependent results. In addition, we find
that while classic vortex identification diagnostics can suggest vortices in the neighborhood of these momentum-blocking vortices, this similarity is without predictive value and fails to generate surfaces
that sufficiently block momentum transport. We believe that the underlying momentum-blocking structures are influencing the gradient-based diagnostics, but the classic diagnostics cannot be used as a proxy measure as they maintain no predictive physical (transport-limiting) value.

The same barrier field trajectory data can also be used to compute TSE fields and identify wall-parallel quasi-planar momentum
transport barrier (MTB) interfaces that locally minimize the 
diffusive transport of momentum. With TRA and TSE fields, the diffusive linear momentum barrier vector field (\ref{eq:instant barrier}), therefore, links vortex
diagnostics with internal fluid interfaces, as seen in tracer field observations, and are objective analogues
of the broadly used but frame-dependent UMZ interfaces and TNTIs.
We have found that MTB interfaces significantly outperform the classic
UMZ interface approaches, in terms of their ability to block viscous
momentum transport, regardless of the specific velocity chosen. The use of TSE calculations for the
barrier equations eliminates sensitivities in UMZ analysis with respect
to changes in the number of bins used in velocity histograms and variations
in the size of the domain of analysis. In comparison, the only free parameter in our vortex and MTB algorithm is the spatial resolution of TSE fields, although this is simply a question of computational resources. Whereas increases in spatial resolution reduce momentum transport through vortex and MTB extractions, increases in the number of velocity histogram bins eventually lead to the disappearance of any UMZ peak or interface.

This first-principles approach of starting with an objective momentum flux provides universal diagnostics based on physics for future structure extraction techniques, the $\Psi$ and $\Psi_N$ fields. Furthermore, the objective nature of this work provides frame-indifferent structures that are experimentally verifiable \citep[e.g.][]{Eisma2021}, unlike the predictions from non-objective criteria. The vortices and MTB identified are resilient under fluid advection, providing a new way in which to track the evolution of individual or clusters of structures. In this way, we hope to provide a common test to verify models and settle longstanding debates in fluids, such as the prevalence or importance of hairpin structures in the boundary layer \citep[see, e.g.,][]{Marusic2019}, and the role of inner versus outer layer motions. 

Techniques for three-component PIV measurements are rapidly advancing \citep{Raffel2018}, and the momentum-transport barrier techniques examined here provide a new way to identify such barriers in an observer-independent fashion in the same flow volume as these data increase in ubiquity. For experimental flows in which volumes of 3D measurements are not available, there is also potential for developing symmetry arguments, such as those provided by \citet{Haller2020}. These can, for instance, simplify the barrier vector field to fewer components for flows with strong anisotropy. Qualitative comparisons between structures in numerical simulations and tracer experiments may also provide insights by determining the roles of analogous features when experimental velocity measurements are unavailable.

Future research should investigate the role that
parabolic invariant manifolds of the momentum barrier equation (\ref{eq:instant barrier})
play in near-wall turbulence as well. \cite{Haller2020} also derive a barrier
equation for the transport of vorticity, which takes the form $\mathbf{x}^{\prime}=\Delta\boldsymbol{\omega}$.
The streamsurfaces of the latter barrier equation generally differ
from those of equation (\ref{eq:instant barrier}) and hence would highlight
different internal interfaces in near-wall turbulence in a theory
that seeks vorticity transport minimizing interfaces. Another extension
of the present theory could use the Lagrangian active barriers introduced
in \cite{Haller2020}, which are material surfaces minimizing momentum
or vorticity transport over a whole time interval rather than just
instantaneously. Such material barriers should mimic the structures
seen in the classic smoke experiments of \citet{Falco1977} even more
closely.

\section*{Acknowledgements}

The Johns Hopkins Turbulence Database used in this study is accessible
at https://doi.org/10.7281/T10K26QW. The authors acknowledge financial
support from Priority Program SPP 1881 (Turbulent Superstructures)
of the German National Science Foundation (DFG), and the Swiss National Science Foundation (SNSF) Postdoc Mobility Fellowship Project P400P2 199190. Declaration of Interests: The authors report no conflict of interest.\bibliographystyle{jfm}

\section*{Appendix: Total Objective Momentum Flux}

\begin{figure}
  \centerline{\includegraphics[scale=0.15]{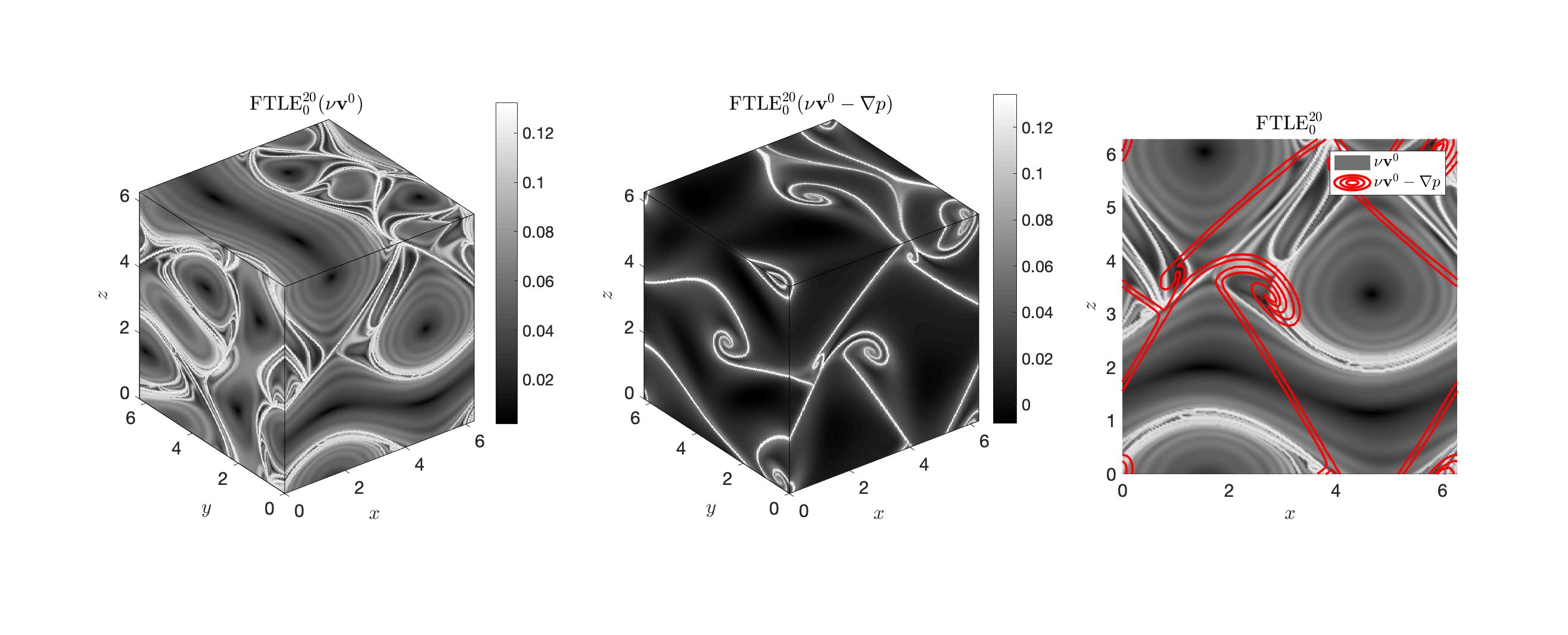}}
  \caption{Comparison of structurally stable transport barriers as identified by FTLE. L to R: Barriers to diffusive momentum flux. Barriers to total momentum flux after including the pressure gradient. Comparison of diffusive and total momentum barriers on the $y=2\pi$ plane.}
\label{fig:ABC_Pressure}
\end{figure}

For the unsteady ABC flow
\begin{equation}
\mathbf{v}(\mathbf{x},t)=e^{-\nu t}\mathbf{v}^{0}\left(\mathbf{x}\right),\qquad\mathbf{v}^{0}\left(\mathbf{x}\right)=\left(\begin{array}{c}
A\sin x_{3}+C\cos x_{2}\\
B\sin x_{1}+A\cos x_{3}\\
C\sin x_{2}+B\cos x_{1}
\end{array}\right),\label{eq:time-dependent ABC flow}
\end{equation}
Theorem 7.5 of \citet{Haller2020} guarantees that the instantaneous barriers to diffusive momentum transport coincide with the structurally stable 2-D invariant manifolds of the flow generated by the steady velocity field $\mathbf{v}^0$, as well as the material and instantaneous barriers to vorticity transport. Specifically, the Eulerian barrier equation for momentum transport
is
\[
\mathbf{x}^{\prime}=\nu\rho\Delta\mathbf{v}=-\nu\rho k^{2}\alpha(t)\mathbf{v}^{0}(\mathbf{x}).
\]
 Note that 
\[
\mathbf{v}_{t}=-\nu\mathbf{v}^{0}\left(\mathbf{x}\right),\quad\nu\Delta\mathbf{v}=-\nu\mathbf{v}^{0}\left(\mathbf{x}\right),
\]
 where the second identity follows from the Beltrami property with
$k=1$ in this case. Therefore, using the Navier\textendash Stokes
equations, we obtain

\begin{eqnarray}
-\frac{1}{\rho}\boldsymbol{\nabla}p=\left(\boldsymbol{\nabla}\mathbf{v}\right)\mathbf{v}&&=\left(\begin{array}{ccc}
0 & -C\sin x_{2} & A\cos x_{3}\\
B\cos x_{1} & 0 & -A\sin x_{3}\\
-B\sin x_{1} & C\cos x_{2} & 0
\end{array}\right)\left(\begin{array}{c}
A\sin x_{3}+C\cos x_{2}\\
B\sin x_{1}+A\cos x_{3}\\
C\sin x_{2}+B\cos x_{1}
\end{array}\right)\nonumber\\ 
&&= \left(\begin{array}{c}
-BC\sin x_{1}\sin x_{2}+AB\cos x_{1}\cos x_{3}\\
BC\cos x_{1}\cos x_{2}-AC\sin x_{2}\sin x_{3}\\
-AB\sin x_{1}\sin x_{3}+AC\cos x_{2}\cos x_{3}
\end{array}\right)
\label{eq:PressureABC}
\end{eqnarray}
The pressure gradient, therefore, has a different magnitude and orientation relative to $\Delta\mathbf{v}=\mathbf{v}^{0}$ . This discrepancy can indeed be seen in Figure \ref{fig:ABC_Pressure} where the FTLE field is calculated for $\Delta\mathbf{v}$ and (\ref{eq:PressureABC}).

We note that for the turbulent channel flow of focus here, the streamwise velocity gradient is predominantly orthogonal to the pressure gradient. Thus, the shear-interfaces used to delineate uniform momentum zones actually maximize the total momentum transport through them.

Following these discussions and the connections drawn between momentum zones and experimentally observable barriers \citep{Westerweel2009}, we find that our definition of active transport that focuses on surfaces
that block the viscous transport (as opposed to the total transport) of momentum and vorticity is the correct one for the purposes of structure identification that is experimentally reproducible.

\bibliography{references}

\end{document}